\documentclass[twocolumn]{aastex631}
\usepackage{CJK}

\usepackage{graphicx}
\usepackage{multirow}
\usepackage{comment}

\newcommand{\Oiii}{[\ion{O}{3}]}
\newcommand{\Oiirec}{\ion{O}{2}}
\newcommand{\Oii}{[\ion{O}{2}]}
\newcommand{\Nii}{[\ion{N}{2}]}
\newcommand{\Niii}{[\ion{N}{3}]}
\newcommand{\Cii}{[\ion{C}{2}]}
\newcommand{\Sii}{[\ion{S}{2}]}
\newcommand{\Siii}{[\ion{S}{3}]}
\newcommand{\Neiii}{[\ion{Ne}{3}]}
\newcommand{\Cliii}{[\ion{Cl}{3}]}
\newcommand{\Ariii}{[\ion{Ar}{3}]}
\newcommand{\Ariv}{[\ion{Ar}{4}]}
\newcommand{\Hii}{\ion{H}{2}}
\newcommand{\Hei}{\ion{He}{1}}

\newcommand{\secpoint}{\mbox{$''\mskip-7.6mu.\,$}}

\newcommand{\um}{\ensuremath{\mu\mathrm{m}}}
\newcommand{\chb}{\ensuremath{c(\mathrm{H}\beta)}}
\newcommand{\wabs}{\ensuremath{W_\mathrm{abs}}}

\newcommand*\diff{\mathop{}\!\mathrm{d}}

\defcitealias{chen23}{C23}

\begin{document}
\begin{CJK*}{UTF8}{gbsn}

\title{Joint Optical and Infrared Observations of N and O Reveal the Dust-Obscured Gas in Haro 3}

\correspondingauthor{Yuguang Chen}
\email{yuguangchen@cuhk.edu.hk}

\author[0000-0003-4520-5395]{Yuguang Chen (陈昱光)}
\affiliation{Department of Physics, The Chinese University of Hong Kong, Shatin, N.T., Hong Kong SAR, China}
\affiliation{Department of Physics and Astronomy, University of California, Davis, 1 Shields Ave., Davis, CA 95616, USA}

\author[0000-0001-5860-3419]{Tucker Jones}
\affiliation{Department of Physics and Astronomy, University of California, Davis, 1 Shields Ave., Davis, CA 95616, USA}

\author[0000-0003-4792-9119]{Ryan L. Sanders}
\affiliation{Department of Physics and Astronomy, University of Kentucky, 505 Rose Street, Lexington, KY 40506, USA}

\author[0000-0002-3698-7076]{Dario Fadda}
\affiliation{Space Telescope Science Institute, 3700 San Martin Drive, Baltimore, MD 21218, USA}

\author[0000-0002-9183-8102]{Jessica Sutter}
\affiliation{Whitman College, 345 Boyer Avenue, Walla Walla, WA 99362, USA}

\author[0000-0002-1261-6641]{Robert Minchin}
\affiliation{National Radio Astronomy Observatory, P.O. Box O, Socorro, NM 87801, USA}

\author[0000-0001-5847-7934]{Nikolaus Z. Prusinski}
\affiliation{Cahill Center for Astronomy and Astrophysics, California Institute of Technology, 1200 E California Blvd, MC 249-17, Pasadena, CA 91125, USA}

\author[0009-0007-0184-8176]{Sunny Rhoades}
\affiliation{Department of Physics and Astronomy, University of California, Davis, 1 Shields Ave., Davis, CA 95616, USA}

\author[0000-0002-2645-679X]{Keerthi Vasan GC}
\affiliation{Department of Physics and Astronomy, University of California, Davis, 1 Shields Ave., Davis, CA 95616, USA}
\affiliation{The Observatories of the Carnegie Institution for Science, 813 Santa Barbara Street, Pasadena, CA 91101, USA}

\author[0000-0002-4834-7260]{Charles C. Steidel}
\affiliation{Cahill Center for Astronomy and Astrophysics, California Institute of Technology, 1200 E California Blvd, MC 249-17, Pasadena, CA 91125, USA}

\author[0009-0002-6927-5259]{Erin Huntzinger}
\affiliation{Department of Physics and Astronomy, University of California, Davis, 1 Shields Ave., Davis, CA 95616, USA}

\author[0009-0009-4213-3630]{Paige Kelly}
\affiliation{Department of Physics and Astronomy, University of California, Davis, 1 Shields Ave., Davis, CA 95616, USA}

\author[0000-0002-4153-053X]{Danielle A. Berg}
\affiliation{Department of Astronomy, The University of Texas at Austin, Austin, TX 78712, USA}

\author[0000-0002-5068-9833]{Fabio Bresolin}
\affiliation{Institute for Astronomy, University of Hawaii, 2680 Woodlawn Drive, Honolulu, HI 96822, USA}

\author[0000-0002-2775-0595]{Rodrigo Herrera-Camus}
\affiliation{Departamento de Astronom\'ia, Universidad de Concepci\'on, Concepci\'on, Chile}
\affiliation{Millennium Nucleus for Galaxies, Concepci\'on, Chile}

\author[0000-0001-9719-4080]{Ryan J. Rickards Vaught}
\affiliation{Space Telescope Science Institute, 3700 San Martin Drive, Baltimore, MD 21218, USA}

\author[0000-0002-4140-1367]{Guido Roberts-Borsani}
\affiliation{University College London, Gower Street, London WC1E 6AE, UK}

\author[0000-0002-9132-6561]{Peter Senchyna}
\affiliation{Carnegie Observatories, 813 Santa Barbara Street, Pasadena, CA 91101, USA}

\author[0000-0003-3256-5615]{Justin S. Spilker}
\affiliation{Department of Physics and Astronomy and George P. and Cynthia Woods Mitchell Institute for Fundamental Physics and Astronomy, Texas A\&M University, 4242 TAMU, College Station, TX 77843-4242, USA}

\author[0000-0001-6106-5172]{Daniel P. Stark}
\affiliation{Department of Astronomy, University of California, 501 Campbell Hall \#3411, Berkeley, CA 94720, USA}

\author[0000-0001-6065-7483]{Benjamin Weiner}
\affiliation{MMT/Steward Observatory, University of Arizona, 933 N. Cherry Street, Tucson, AZ 85721, USA}

\author[0000-0002-8650-1644]{D. Christopher Martin}
\affiliation{Cahill Center for Astronomy and Astrophysics, California Institute of Technology, 1200 E California Blvd, MC 249-17, Pasadena, CA 91125, USA}

\author[0000-0003-2821-1750]{Mateusz Matuszewski}
\affiliation{Cahill Center for Astronomy and Astrophysics, California Institute of Technology, 1200 E California Blvd, MC 249-17, Pasadena, CA 91125, USA}

\author[0000-0003-2064-4105]{Rosalie C. McGurk}
\affiliation{W.M. Keck Observatory, 65-1120 Mamalahoa Hwy, Waimea, HI, USA}

\author[0000-0002-0466-1119]{James D. Neill}
\affiliation{Cahill Center for Astronomy and Astrophysics, California Institute of Technology, 1200 E California Blvd, MC 249-17, Pasadena, CA 91125, USA}

\begin{abstract}
Accurate chemical compositions of star-forming regions are a critical diagnostic tool to characterize the star formation history and gas flows which regulate galaxy formation. However, the abundance discrepancy factor (ADF) between measurements from the ``direct'' optical electron temperature ($T_e$) method and from the recombination lines (RL) represents $\sim0.2$ dex systematic uncertainty in oxygen abundance. The degree of uncertainty for other elements is unknown. We conduct a comprehensive analysis of O$^{++}$ and N$^+$ ion abundances using optical and far-infrared spectra of a star-forming region within the nearby dwarf galaxy Haro 3, which exhibits a typical ADF. Assuming homogeneous conditions, the far-IR emission indicates an O abundance which is higher than the $T_e$ method and consistent with the RL value, as would be expected from temperature fluctuations, whereas the {far-IR} N abundance is too large to be explained by temperature fluctuations. A two-phase analytical model reveals that differential dust obscuration associated with temperature inhomogeneity is likely required to explain all the emission line ratios, and that the total oxygen metallicity of two phases is consistent with the RL metallicity.  Our findings underscore the critical importance of resolving the cause of abundance discrepancies and understanding the biases between different metallicity methods. This work represents a promising methodology, and we identify further approaches to address the current dominant uncertainties. 
\end{abstract}


\section{Introduction}
\label{sec:intro}

The abundance of heavy elements in the interstellar medium (ISM) with respect to hydrogen, i.e., metallicity, serves as a critical diagnostic tool for understanding the astrophysical processes which govern galaxy formation. Metallicity is particularly valuable as a probe of star formation history, gas accretion, feedback mechanisms, and galactic-scale outflows \citep[e.g.,][]{maiolino2019,sanders2021,dave2017,torrey2019}. Oxygen (O) and nitrogen (N) are among the elements most commonly used to assess gas-phase metallicities, and they are more powerful in combination due to their distinct nucleosynthetic origins and subsequent enrichment pathways.
Oxygen in the ISM is produced mostly from the cataclysmic termination of massive stars in the form of core-collapse supernovae (CCSNe). In contrast, nitrogen's enrichment is largely facilitated by stellar winds from intermediate-mass Asymptotic Giant Branch (AGB) stars \citep[e.g.,][]{nomoto13}, resulting in a temporally delayed contribution to the enrichment of the ISM relative to star-forming events. The total nitrogen yield also increases with overall metallicity due to secondary production from the CNO cycle. 
This dichotomy in enrichment processes renders the ratio of these elements an invaluable metric for tracing the integrated history of star formation and gas flows.

Substantial research has been dedicated to understanding the enrichment patterns of O and N metallicities \citep[e.g.,][]{alloin79, pilyugin12, berg12, izotov12, james15, vincenzo16, steidel16, magrini18, annibali22, sanders23}, revealing minimal evolution in the N/O ratio among low-metallicity galaxies while observing an elevated N/O ratio in high-metallicity galaxies. This trend aligns with the classical CCSNe and AGB enrichment scenarios. Nevertheless, recent findings have uncovered an unexpected increase in the N/O ratio within early galaxies during the epoch of reionization  \citep{bunker23, isobe23, marqueschaves24}, possibly linked to settings of supermassive star enrichment processes operating in dense clusters \citep{dantona23, nagele23, pascale23, belokurov23, charbonnel23, senchyna23, cameron23, kobayashi24, nandal24}. 
Notably, the large N/O abundances found in early galaxies are based on rest-frame ultraviolet emission lines, as opposed to optical diagnostics used at lower redshifts ($z\lesssim3$). Such diagnostics are powerful and complementary in probing a vast redshift range, but may introduce different systematic errors \citep[e.g., ][]{topping24}. 
This underscores the necessity to accurately determine the chemical abundance patterns in galaxies to better understand their chemical evolution throughout cosmic history. 

A standard approach to measure the gas-phase metal content, often referred to as the ``direct'' $T_e$ method, relies on analyzing the collisionally excited emission lines (CEL). These are produced when ions such as O$^{++}$ or N$^+$ undergo radiative transitions from various energy levels, following excitation from collisions with electrons. The flux ratio of emission lines from different collisionally excited energy levels (e.g., \Oiii~$\lambda$4363 / \Oiii~$\lambda$5007) provides a measurement of the temperature $T_e$, which in turn allows ion abundances to be determined from line luminosities. 
However, this $T_e$-sensitivity presents a challenge when multiple ionizing sources contribute to temperature structures and fluctuations within the gas \citep{pilyugin12}. Treating the gas as a uniform phase across an entire galaxy -- or even within an \Hii\ region -- can bias the $T_e$ method toward regions of higher temperature, and thus, lower elemental abundance, leading to an underestimation of the true metallicity. This systematic bias is often invoked to explain the so-called abundance discrepancy factor (ADF), where $T_e$-based oxygen metallicities are consistently lower than those determined through $T_e$-insensitive recombination lines (RL; e.g., \Oiirec\ 4650 V1 multiplet, also emitted by O$^{++}$ gas) by $\sim 0.25$ dex \citep{peimbert93, peimbert05, esteban09, esteban14}. 
As one way to explain this dicrepancy, the amount of bias in the $T_e$ method is directly related to the magnitude of temperature fluctuations, which can be expressed as a dimensionless variance $t^2$ \citep{peimbert67}:
\begin{equation}
\label{eq:temperature_fluctuations}
t^2 = \frac{\int n_X n_e (T_e - T_0)^2 \diff V}{\int T_0^2 n_X n_e \diff V }
\end{equation}
where $n_X$ and $n_e$ are the densities of specific ions and free electrons respectively, $T_0$ is the mean electron temperature, and $V$ is the integration volume. 
A key open question is whether temperature fluctuations in ionized nebulae are in fact sufficiently high to explain the ADF \citep[e.g.,][]{peimbert17,maiolino2019}. 

In a pilot study by \citet{chen23} (hereafter \citetalias{chen23}), the temperature fluctuation hypothesis was empirically tested within the nearby dwarf galaxy Mrk~71 using far-IR fine-structure emission lines. While these far-IR emission lines (e.g., \Oiii~$\lambda\lambda$52, 88~\um) are the product of collisional excitation, the low energy required for their excitation renders them effectively insensitive to $T_e$. Therefore, a direct comparison between the far-IR-derived metallicity and the optical-CEL (OCEL) metallicity yields a direct gauge of temperature fluctuations without ambiguity caused by different emission mechanisms \citep{jones20}, i.e., the metallicity derived from far-IR lines is expected to be nearly identical to the RL method even if $t^2$ is large.
Intriguingly, our findings revealed a $t^2$ value in Mrk~71 for O$^{++}$ gas that is consistent with zero, and is in tension ($\sim 2\sigma$) with the ADF derived between the O$^{++}$ RL and OCEL metallicities. 
This result highlights the power of combining optical and far-IR spectra to address key issues regarding the magnitude of temperature fluctuations $t^2$, the origin of the ADF, and the absolute gas-phase abundances. Such methods are especially promising for application at high redshifts, where the relevant optical and far-IR lines are now being detected at $z>8$ thanks to the combination of JWST and the Atacama Large Millimeter Array (ALMA; \citealt{fujimoto23,jones20, usui25, harikane25}). These high-$z$ observations suggest that a ``two-zone'' model in temperature and density is required to explain the observed discrepancy between the optical and far-IR emission fluxes, and that a significant fraction of high-density gas is hidden from the far-IR emission due to the low critical densities of the far-IR emission lines \citep[e.g.,][]{harikane25}. It is thus a pivotal time to harmonize metallicity measurements across both low and high-redshift galaxies.

Characterization of the temperature fluctuations within ionized nebulae can be significantly expanded by using the same methods to measure $t^2$ between different ion species. A recent study by \citet{md23} underscores the necessity by unveiling a correlation between $t^2(\mathrm{O}^{++})$ and the differential temperature, $T_e(\mathrm{O}^{++}) - T_e(\mathrm{N}^+)$. This suggests that temperature fluctuations might be intrinsically linked to the inherent temperature disparities amongst various gas phases. 

In this paper, we apply the optical+far-IR methodology from \citetalias{chen23} to measure $t^2$ from both O$^{++}$ as well as N$^+$ for the first time. This study is made possible by the recent availability of comprehensive datasets from integral field unit (IFU) spectrographs spanning the optical and far-IR: the Far Infrared Field-Imaging Line Spectrometer on the Stratospheric Observatory for Infrared Astronomy (SOFIA/FIFI-LS; \citealt{fischer18}), the Photodetector Array Camera and Spectrometer on the Herschel Space Telescope (Herschel/PACS; \citealt{poglitsch10}), and the recent commissioning of the red channel (also known as the Keck Cosmic Reionization Mapper; KCRM) of the Keck Cosmic Web Imager on the W.M. Keck Observatory (Keck/KCWI; \citealt{morrissey18}). 
Here we focus on a star-forming region within the dwarf galaxy Haro~3 (also known as NGC~3353 or Mrk~35; RA=10:45:22.4; Dec=$+$55:57:37; $z=0.003208$). Based on the fiber spectrum centered on the host galaxy from the Sloan Digital Sky Survey data release 8 (SDSS DR8), the host galaxy has a stellar mass $\log(M_*/M_\odot) \sim 8.5$, star formation rate $\mathrm{SFR}\sim 0.1 M_\odot / \mathrm{yr}$. {The gas-phase metallicity was measured as $12 + \log(\mathrm{O}/\mathrm{H}) = 8.296 \pm 0.013$ from long-slit spectrum using the direct-$T_e$ method by \citet{izotov04}, while the SDSS measured $12 + \log(\mathrm{O}/\mathrm{H}) \simeq 8.7$ \citep{kauffmann03, brinchmann04, tremonti04} for the host galaxy, although we note that metallicities adopted from SDSS are systematically higher than from the $T_e$ method \citep[e.g.,][]{kewley08}}. In this work, we focus on the luminous \Hii\ region located $\simeq 4\secpoint2$ ($\sim 300$ pc) northwest of Haro~3's nucleus, often labeled as ``Region A'' \citep{steel96}. This region dominates the host galaxy's total star formation and ionizing photon production \citep{johnson04}. 

This paper is organized as follows. We describe the observational dataset in \S2. Detailed methods leading to the temperature and metallicity measurements are presented in \S3. In \S4, we discuss our findings. The conclusions are summarized in \S5.  

\section{Observations}

The Keck/KCWI data of Haro~3 were collected on the nights of UT 2023-06-19 and 2023-11-11. Both nights were dark, clear, with photometric conditions. The observations involved two distinct observation strategies: firstly, a series of brief snapshots to prevent the saturation of strong emission lines, and secondly, deep exposures aimed at detecting faint emission features including auroral and oxygen recombination lines. All observations used the Medium slicer, with 0\farcs7 slits and a $\sim 16" \times 20"$ field of view (FoV). For the snapshots, the BL and RL gratings were employed to obtain broad wavelength coverage. The BL grating was centered at 4500~\AA\ (coverage 3580--5580 \AA, $R \sim 1800$), with an exposure time of $t_\mathrm{exp} = 5\mathrm{s} \times 8 = 40$s; the RL grating was centered at 8000~\AA\ (6465--9780 \AA, $R\sim 2000$) with $t_\mathrm{exp} = 30\mathrm{s}\times 7 = 210$s. The deep exposures utilized the BM and RL gratings. The BM grating was centered at 4400~\AA\ (3985--4865 \AA, $R \sim 4000$) to resolve the \Oiirec\ RL multiplet with individual exposure times ranging between 200s, 600s, and 660s, and a total $t_\mathrm{exp} = 2660$s; the RL grating was centered at 7250~\AA\ (5715--9035 \AA, $R \sim 1900$) to cover the faint \Nii\ auroral emission, with individual exposure times as a mixture of 200s, 300s, and 330s, achieving a total $t_\mathrm{exp}=2660$s as well. Between individual exposures, the FoV was rotated by $45^\circ$ to enable good spatial sampling. 

\begin{figure*}[ht!]
\includegraphics[width=0.38\textwidth]{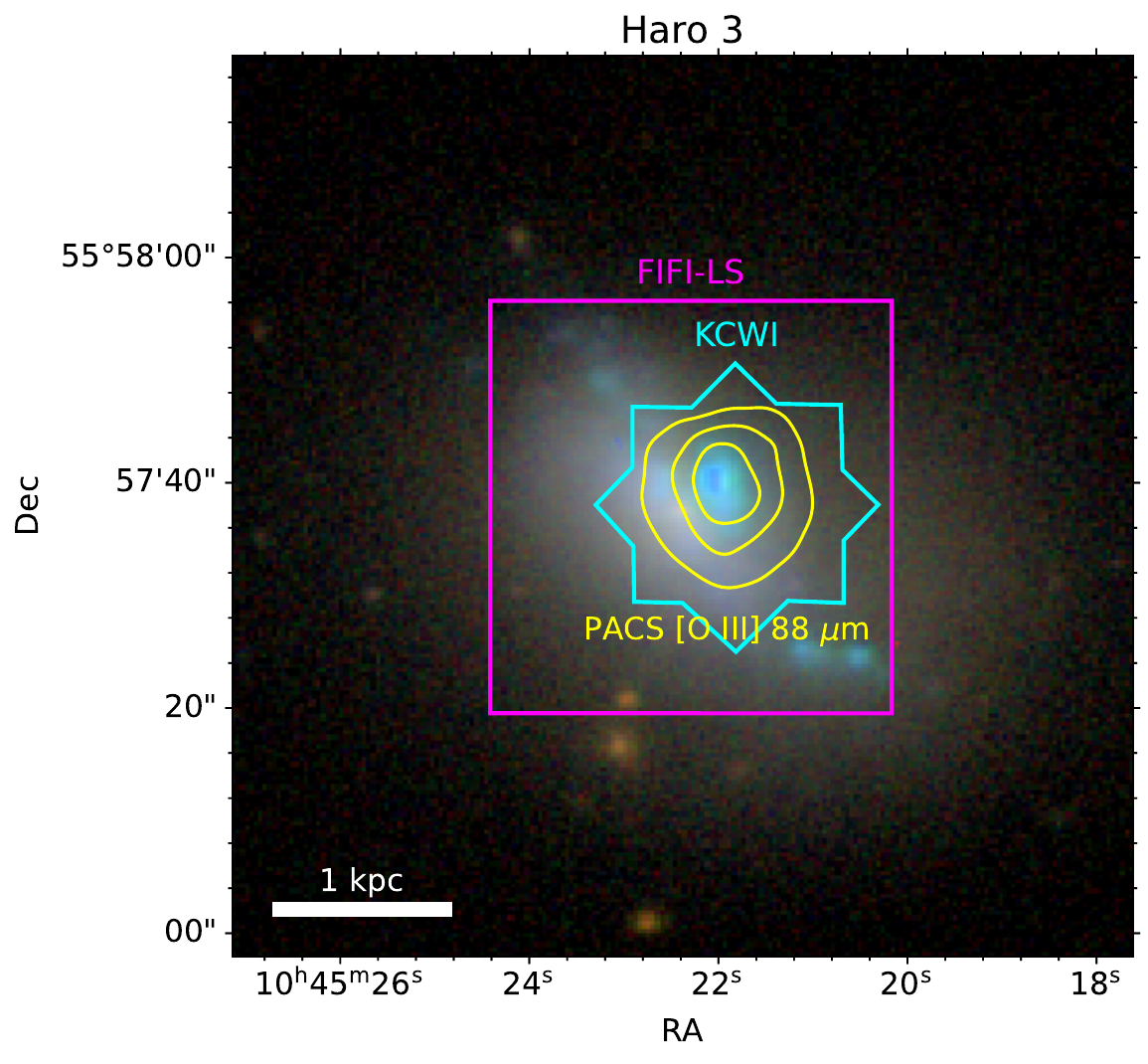}
\includegraphics[width=0.61\textwidth]{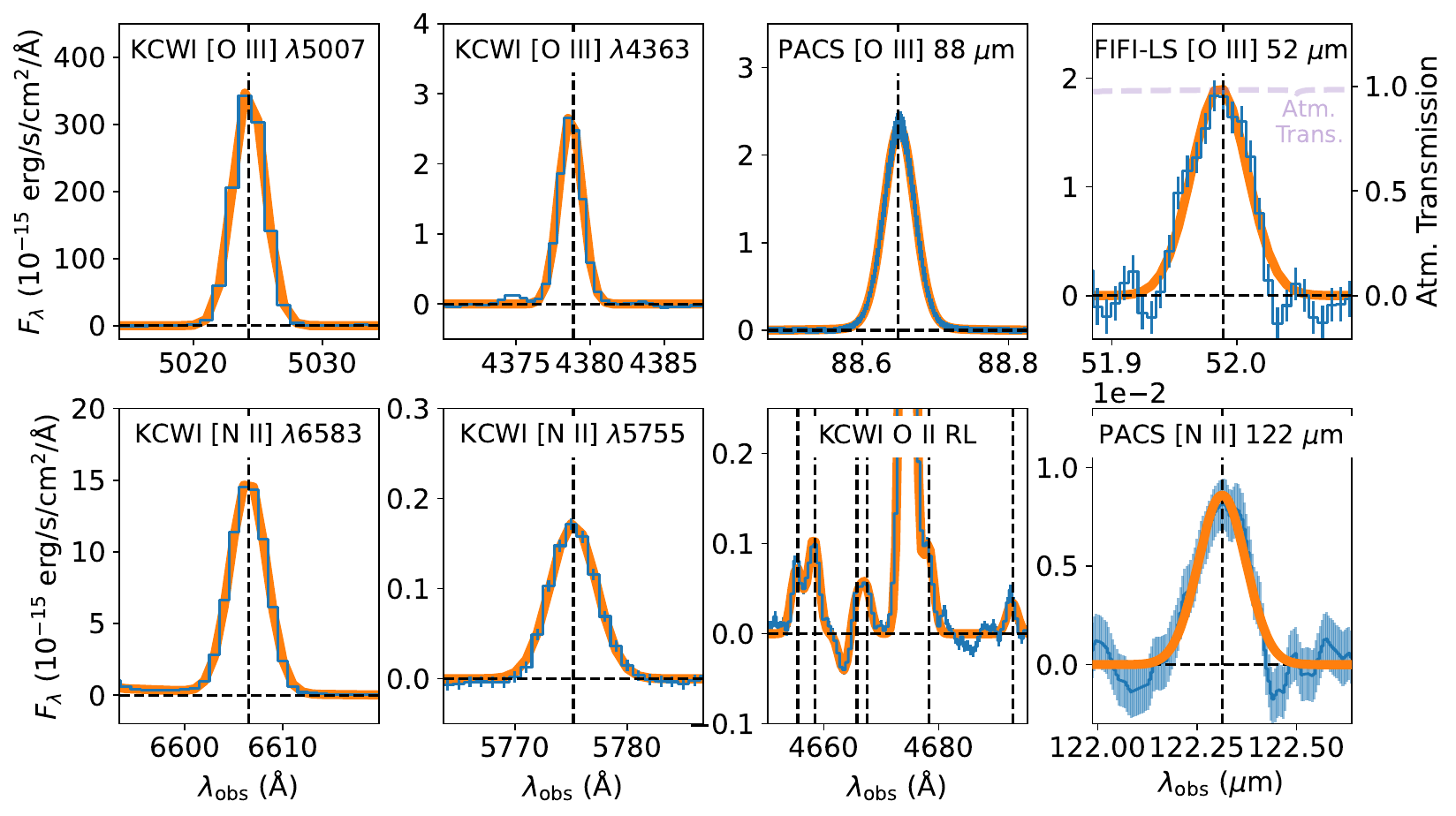}\\
\includegraphics[width=0.99\textwidth]{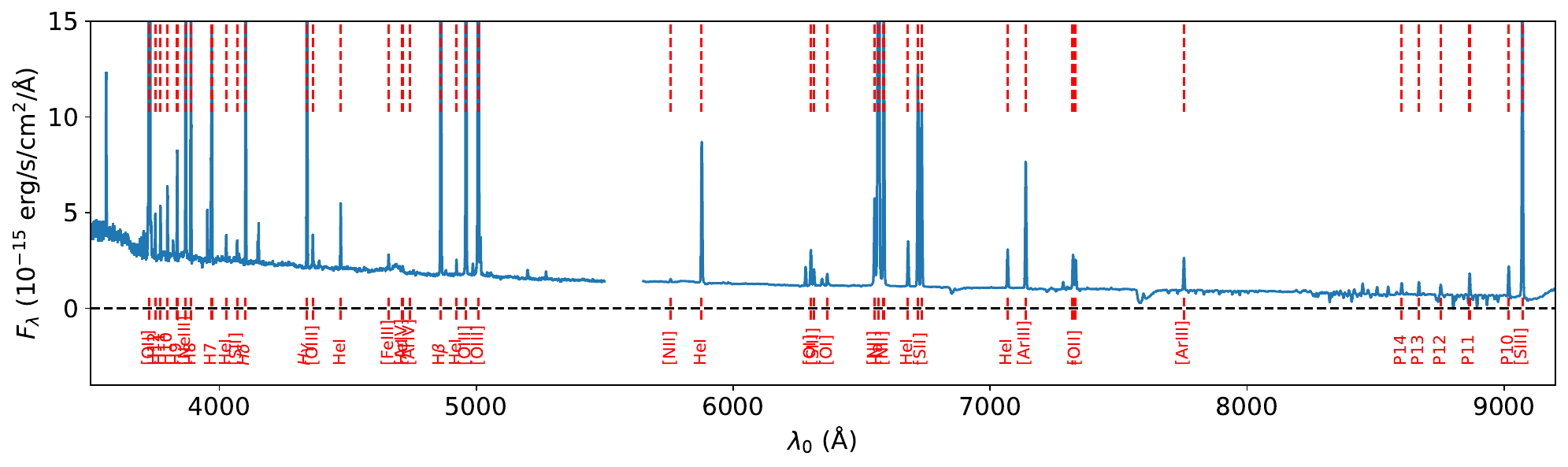}
\caption{ \textit{Top-left:} Fields of view of the Keck/KCWI (cyan) and SOFIA/FIFI-LS (magenta) observations, and contours of the \Oiii\ 88~\um\ emission map from Herschel/PACS (yellow), overlaid on the color composite image constructed from Sloan Digital Sky Survey images in \textit{g}, \textit{r} and \textit{i} filters. \textit{Top-right:} Extracted spectra after continuum subtraction (blue) and the corresponding fitted line profiles (orange) of several important emission features for this work. {\textit{Bottom:} The full Keck/KCWI optical spectrum of Haro~3 in the 2\secpoint25 aperture observed from the BL-4500 and RL-7250 grating setups. A subset of prominent emission features are marked in red.}
\label{fig:fov}}
\end{figure*}

We performed the KCWI data reduction process following the methodology described in \citetalias{chen23}, utilizing the upgraded Python version of the Data Reduction Pipeline (DRP)\footnote{\url{https://github.com/Keck-DataReductionPipelines/KCWI_DRP}}, which now supports the red channel. As in \citetalias{chen23}, we altered the DRP, opting for linear resampling over cubic resampling to ensure flux conservation. Sky subtraction was performed by scaling and removing the b-spline sky model generated by the DRP from separately obtained exposures of adjacent blank sky, subsequent to the on-source exposures. Flux calibration was performed using observations of the A-type standard star BD+26~2606, taken immediately after the science exposure sequence. Data cubes from individual exposures were resampled to a uniform grid with a pixel size of $0\secpoint3 \times 0\secpoint3$ prior to their combination using \textsc{KcwiKit} \citep{kcwikit}. The blue-channel data were averaged with weights according to exposure time. The red-channel detector exhibited a much higher incidence of cosmic rays (CR), which were not entirely eliminated by the DRP. To address this, we employed a sigma-clipping method with $N_\sigma = 3$, centered around the median value, prior to the average combination of data cubes. From comparison of flux measurements across different grating configurations, we determined that the relative flux accuracy is within $<2\%$. Additionally, two pseudo-broad-band images were generated from the RL-7250 and RL-8000 cubes in the \textit{i} band and compared with images from the SDSS data release 18 \citep{almeida23}. A discrepancy of $\sim4\%$ was identified and subsequently corrected in the data cubes. Following this adjustment, we conservatively adopt a {5\%} absolute flux accuracy, which is included in the subsequent analyses. For close doublets, e.g., \Oii\ $\lambda\lambda 3726,29$ and \Sii\ $\lambda\lambda 6716,31$, the relative flux calibration uncertainty should be negligible. However, we conservatively impose a minimum 1\% flux error, resulting in an increase of the $n_e$(\Sii) uncertainty from $\sim 10~\mathrm{cm}^{-3}$ to $\sim 20~\mathrm{cm}^{-3}$.
The reported uncertainties of those line ratios only come from the random noise. 

The \Oiii~$\lambda$88~\um\ and \Nii~$\lambda$122~\um\ emission lines were observed with the Photodetector Array Camera and Spectrometer (PACS; \citealt{poglitsch10}) aboard the Herschel Space Observatory, as part of the Herschel Dwarf Galaxy Survey (HDGS; \citealt{cormier15}). These reduced data (v14.2.0) were obtained from the Herschel Data Archive\footnote{\url{https://archives.esac.esa.int/hsa/whsa/}}. The``chop-nod'' mode was employed for these observations, with exposure times of 4068~s for \Oiii88\um\ and 805~s for \Nii122\um. Additionally, reduced data of \Oiii~52~\um\ emission were acquired from the SOFIA Data Archive\footnote{\url{https://irsa.ipac.caltech.edu/Missions/sofia.html}}, as reported by \citet{peng21}. These observations utilized the Field-Imaging Far-Infrared Line Spectrometer (FIFI-LS; \citealt{fischer18}), with a total integration time of 1413~s. The flux calibration uncertainty for the PACS and FIFI-LS data represents one of the primary limiting factors in this analysis. For both PACS and FIFI-LS, flux calibration is based on the telescope internal thermal background and/or solar-system bodies (planets, moons, and asteroids) used as flux standards. The absolute flux calibration uncertainty for PACS is $\simeq 11$--12\%\footnote{PAC Observer's Manual: \url{http://herschel.esac.esa.int/Docs/PACS/html/ch04s10.html}}, established from observations of $\sim$30 absolute flux calibration sources. \citet{fadda23} compared flux measurements of sources observed by both PACS and FIFI-LS and found an RMS scatter of $\simeq 2\%$ for the \Cii\ $\lambda 158~\um$ emission and 3\% for the \Oiii\ $\lambda 88~\um$ emission. In \citet{chen23}, we compared the \Cii\ $158~\um$ emission in Mrk~71 measured with PACS and FIFI-LS and found that the flux difference is $<10\%$. We therefore adopt a uniform 11\% absolute flux calibration uncertainty for all PACS- and FIFI-LS-measured fluxes, which propagates to a 15\% uncertainty in the far-IR flux ratios.

\section{Methods}

\subsection{Flux Measurements\label{sec:flux_measurements}}
\begin{figure*}[ht!]
    \centering
    \begin{minipage}{0.49\textwidth}
        \centering
        (a) PSF Matching
        \includegraphics[width=\linewidth]{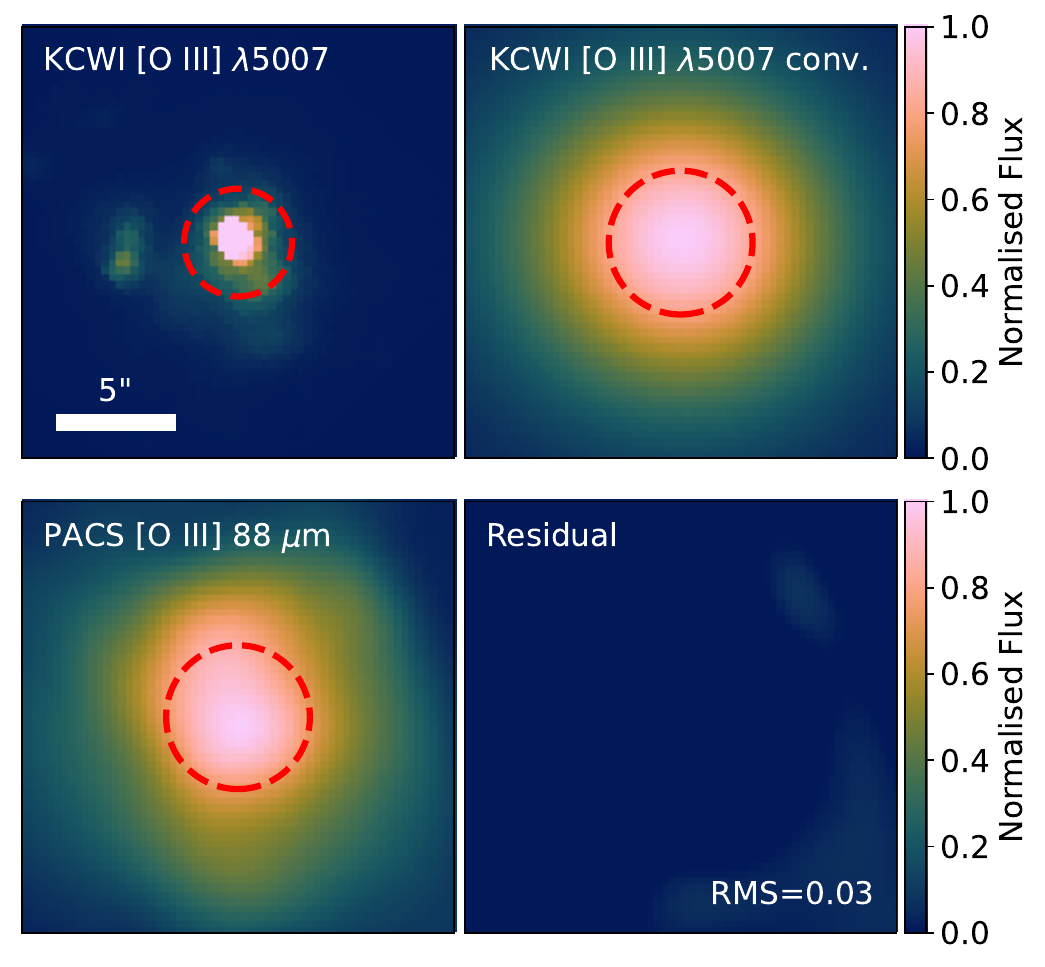}
    \end{minipage}
    \hfill
    \begin{minipage}{0.49\textwidth}
        \centering
        (b) Attenuation Correction
        \includegraphics[width=0.95\linewidth]{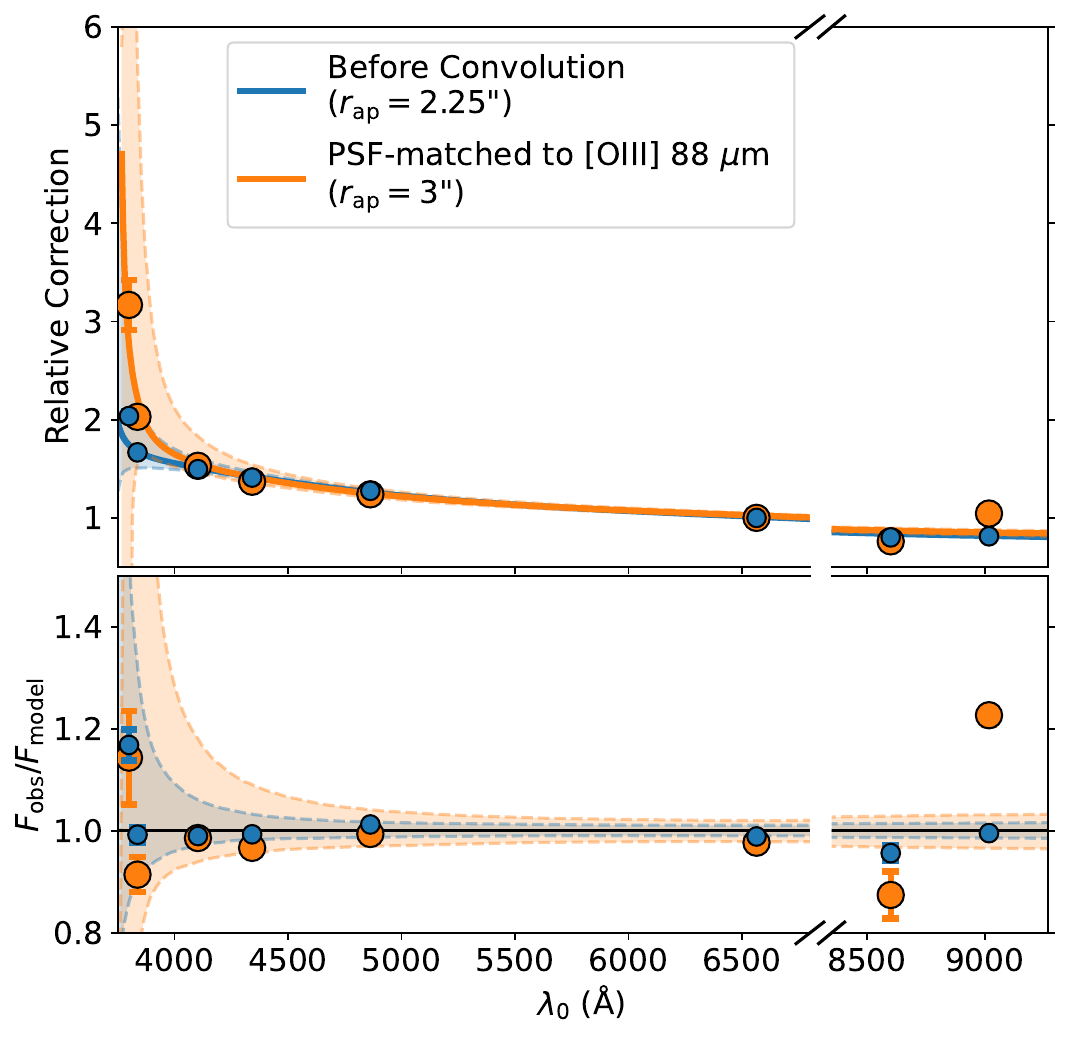}
        \end{minipage}
    \caption{ \textit{Left}: An example of PSF matching between the \Oiii\ $\lambda5007$ PNB image from Keck/KCWI (top-left) and the \Oiii\ 88~\um\ PNB image from Herschel/PACS (bottom-left).  The KCWI PNB image has been convolved with a 2D Gaussian kernel (top-right) to match the PACS PNB image's PSF. The resulting residual is shown in the bottom-right panel. The red dashed circles indicate the apertures from which the 1D spectra were extracted. 
    \textit{Right}: Modeling of dust attenuation based on the observed flux ratios of hydrogen Balmer and Paschen lines. In the top panel, the data points represent the correction factors needed to align the observed flux ratios with those predicted by Case B recombination. The best-fit attenuation models are represented by solid curves, while the shaded areas denote their 1$\sigma$ uncertainties. The blue (orange) color indicates the 1D spectra before (after) PSF matching. The steep slope at the lower wavelengths is primarily due to contributions from stellar absorption. The lower panel shows the fractional deviations between the observed and best-fit flux ratios, for the same data points. Data points near unity indicate a good agreement with the attenuation model. 
        \label{fig:methods}}
\end{figure*}

Accurate temperature and metallicity measurements are based on the ability to reliably compare line fluxes from optical and infrared data. 
This comparison requires matching the point spread function (PSF) and extraction aperture of the various IFU data sets. 
We matched the PSF of each instrument by convolving pseudo-narrowband (PNB) images of distinct \Oiii\ or \Nii\ emission lines with Gaussian spatial profiles. Fig.~\ref{fig:methods} shows an example of PSF matching between KCWI \Oiii\ $\lambda5007$ and PACS \Oiii\ $88$~\um. For the \Oiii\ $\lambda$5007 (KCWI) compared to far-IR images, the optimal full width at half maximum (FWHM) values for the Gaussian convolution kernels were determined from the best fit to be $10\secpoint03 \pm 0\secpoint04$ for \Oiii\ $\lambda$52~\um\ (FIFI-LS), $9\secpoint45 \pm 0\secpoint02$ for \Oiii\ $\lambda$88~\um\ (PACS), and $13\secpoint5 \pm 0\secpoint1$ for \Nii\ $\lambda$122~\um\ (PACS). The residual root mean square (RMS) of the fit is $<10\%$ of the peak intensity of the \Oiii\ and \Nii\ PNB images. 
These FWHM values are consistent with expectations from the diffraction limit of the telescopes and wavelengths of the targeted features. 

Given the substantial PSF differences between KCWI (with $\sim$1\arcsec\ FWHM) and the far-IR instruments, coupled with the compact nature of Haro 3, the S/N ratio for faint emission lines in the KCWI data significantly deteriorates post-convolution. To ensure a reliable comparison while preserving an acceptable S/N, we extract 1-D spectra from both pre- and post-PSF-matched KCWI data cubes across various apertures. An aperture-correction factor is then derived by comparing the extinction-corrected fluxes (see \S\ref{sec:attenuation}) of strong emission lines (including \Neiii\ $\lambda 3869$, \Oiii\ $\lambda 4363$, $\lambda\lambda4959, 5007$, and H$\beta$) between these spectra. {This approach assumes that the gas properties, especially the chemical enrichment, is uniform in the nebula, regardless of the variations in dust attenuation. We tested the validity of this assumption by measuring the $O^{++} / H^+$ abundances from the optical CEL emission (see \S\ref{sec:temperature_metallicity} below), and found that the abundance in the larger aperture is lower than that of the smaller aperture by $\simeq 0.05$~dex, indicating a slight metallicity gradient.} The apertures, centered on the nebular emission peak, have pre-convolution radii of 2\secpoint25 and post-convolution radii of 3\arcsec\ (Fig. \ref{fig:methods}, left). The aperture sizes were selected to optimize the S/N ratio and to minimize the inclusion of flux from other sources. The systematic uncertainty of this aperture correction is estimated to be $<2\%$, based on the standard deviation of the flux ratios. 

The spectral profiles of the optical emission lines are best described by a narrow Gaussian profile with an additional broad component contributing $<1\%$ of the total flux. However, due to undersampling in detector space for SOFIA/FIFI-LS and Herschel/PACS, the far-IR line profiles are instead better described by Voigt profiles. 
For consistency, fluxes of all \Oiii\ and \Nii\ optical and far-infrared CELs were measured by fitting Voigt profiles after extracting the 1D spectra, except for the weak features specifically described below. 
The choice of Voigt versus Gaussian profiles results in $<1\%$ differences in flux for the optical lines. The \Nii\ $\lambda\lambda$6548, 6583 doublet, located near H$\alpha$, is fit concurrently with the intensity ratio constrained to the theoretical value of 1:2.942 \citep{fischer04}. {Fitting them independently yields a total flux consistent within 0.6\% of the joint fit.} The \Nii\ $\lambda$5755, $\lambda$122~\um\ lines, and the \Oiirec\ RL V1 multiplet lines are fitted using only Gaussian profiles due to their relative weakness. Additionally, individual lines within the RL multiplet are fitted simultaneously to account for blending. Fig. \ref{fig:fov} shows the best fits to several representative lines. {Similarly, for the \Oii\ $\lambda\lambda3726,29$ doublet and the \Oii\ $\lambda\lambda$7318,20,30,31 multiplet, the lines are fitted simultaneously assuming the same velocity dispersion.} The 1-$\sigma$ flux uncertainties are derived from the error spectra. However, data cube resampling and PSF convolution inevitably induce pixel covariance, affecting the accuracy of error estimation. To address this, we compare the standard deviation of a featureless section of the spectra with the calculated error spectra, and rescale the error spectra based on the factorial differences before propagating the flux errors. {All measured fluxes related to this work are presented in Appendix~\ref{sec:line_fluxes}}.

\subsection{Attenuation Correction \label{sec:attenuation}} 

In this study, dust attenuation correction, including contributions from the Milky Way\footnote{The Milky Way extinction in the direction of Haro 3 contributes only $\chb \simeq 0.01$ based on \citet{schlafly11}.} and Haro 3, was performed using the Hydrogen Balmer and Paschen lines. Unlike the emission lines from O$^{++}$ and N$^+$, fluxes of the \ion{H}{1} lines for attenuation correction are determined by integration over a wavelength range manually selected to accommodate stellar absorption features. This consideration is particularly crucial for the weaker Balmer lines, where stellar absorption has a significant impact. Reliable flux measurements are obtained from H$\beta$ to H10 (or H$\theta$), as well as P10 and P14 lines. However, the H7 (H$\epsilon$) and H8 (H$\zeta$) lines are excluded due to their blending with other spectral features. Other Balmer and Paschen lines are deemed unsuitable for reliable flux measurements, either being too weak or contaminated by telluric OH emission. The ability to measure H$\alpha$, P10, and P14 fluxes from the newly commissioned red-channel of KCWI significantly increases the accuracy of this attenuation correction. With Paschen line attuenuation being $\sim 70\%$ that of the Balmer lines, and the wavelength baseline increasing by $>4\times$, the attenuation correction uncertainty is reduced by $40\%$ compared to using Balmer lines alone. 

\begin{deluxetable}{ccc}
\tablecaption{{Electron densities and dust attenuation of Haro~3 \label{tab:ne}}}
\tablehead{ $n_e$ & $r_\mathrm{ap} = 2\secpoint25$ & $r_\mathrm{ap} = 3"$\\
\colhead{(cm$^{-3}$)} & (before PSF matching) & (after PSF matching)}
\startdata
\Oiii\ & --- &  $117_{-76}^{+53}$ \\
\Oii\ & $163 \pm 42$ & $101 \pm 23$ \\
\Oiirec\ RL & $578_{-141}^{+76}$ & --- \\
\Oiirec\ RL\tablenotemark{a} & $440^{+230}_{-130}$ & --- \\
\Sii\ & $228 \pm 24$ & $145 \pm 22$ \\
\Cliii\ & $390_{-390}^{+540}$ & $590_{-590}^{+870}$ \\
\Ariv\ & $540_{-400}^{+410}$ & $<1120$ \\\hline
\chb\ &  $0.317 \pm 0.016$ & $0.253 \pm 0.037$ \\
\wabs\ (\AA) & $0.42 \pm 0.34$ & $0.98 \pm 0.51$ \\
\hline
\enddata
\tablenotetext{a}{Using only \Oiirec\ $\lambda\lambda$4638.86, 4641.81, and 4676.23 lines. }
\end{deluxetable}

The strength of dust attenuation, \chb, is derived by modeling the observed fluxes of Balmer and Paschen lines. The model assumes that the intrinsic H line ratios are given by the case B recombination scenario with $T_e =$~8,500~K and an electron density $n_e = 100~\mathrm{cm}^{-3}$. \footnote{This assumption is informed by the $T_e$ and $n_e$ values derived from \Oiii\ far-IR emission, as discussed in \S\ref{sec:temperature_metallicity}. Variations in $T_e$ and $n_e$ have a marginal impact on the results. For example, a scenario with $T_e=10,000$~K and $n_e=100~\mathrm{cm}^{-3}$ alters the P13/H$\beta$ ratio by $<2\%$.} The attenuation of H fluxes is calculated using an extinction curve from \citet{cardelli89} with a parameter $R_V = 3.1$. {Changing $R_V$ assumptions imposes a small change in the final metallicity measurements, mostly for the far-IR metallicities. Specifically, assuming $R_V=2.6$(4.1) modifies the far-IR {O$^{++}$/H$^{+}$} metallicity measurement by $\simeq +0.03$($-0.05$) dex, which does not affect our conclusions.} The model also accounts for stellar absorption by assuming a fixed equivalent width for the Balmer lines (\wabs), while positing no absorption for the Paschen lines. This assumption is based on stellar population synthesis models. Specifically, for the Binary Population and Spectral Synthesis (BPASS) v2.3 \citep{stanway18} model, the Paschen \wabs\ is $\lesssim 30\%$ of the Balmer ones in the age range of $10^7$--$10^9$ years. In the observed spectra, the Paschen lines also have no broad absorption signatures that appear in the high-order Balmer lines. {Assuming a $W_\mathrm{abs}$ for the Paschen lines would result in a poor fit ($\sim 30\%$ deviations between the observed and model fluxes)}. {In the end, the effect of $\wabs$ is small. Assuming $\wabs = 0$~\AA\ results in an increase in \chb\ of 0.01, corresponding to an increase of the dust-corrected \Oiii\ $\lambda5007$ flux of 2\%.} We find that both \chb\ and \wabs\ vary significantly between the small unconvolved optical and the large PSF-convolved far-IR apertures. The change is consistent with the \chb\ and \wabs\ maps derived from the same method using the unconvolved KCWI data. The maps indicate that the larger aperture includes more contribution from the surrounding host galaxy, with both higher dust attenuation and stronger effects of stars with Balmer absorption. 
Consequently, \chb\ measurements were made separately both before and after aperture matching, with the aperture correction factor calculated based on the attenuation-corrected fluxes. The best-fit models (\chb\ and \wabs) are presented in Fig. \ref{fig:methods} and Table \ref{tab:ne}. Uncertainties in the \chb\ measurements are included in the subsequent measurements. 

\subsection{Temperature, Density, and Metallicity \label{sec:temperature_metallicity}}

\begin{figure*}[ht!]
\includegraphics[width=0.33\textwidth]{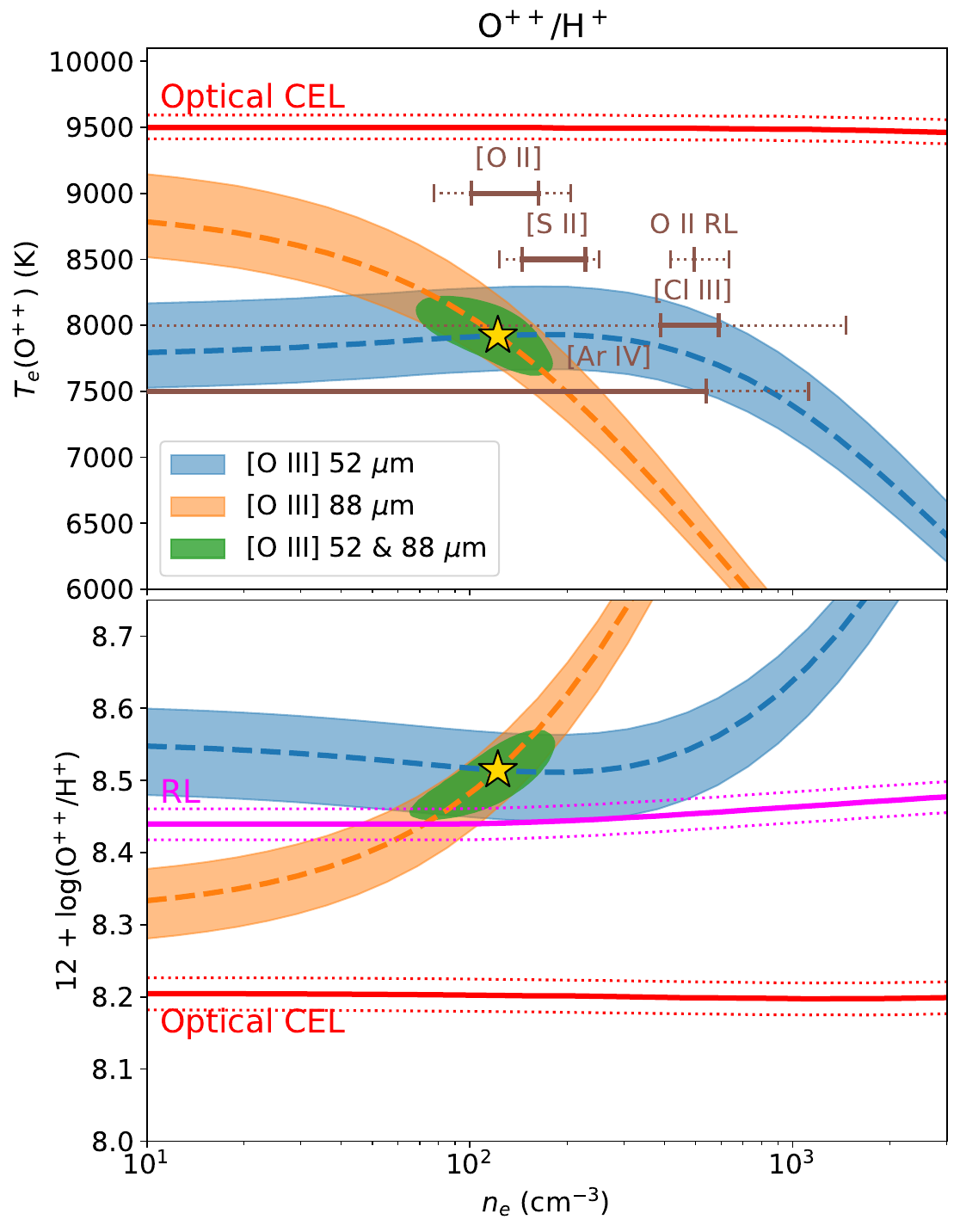}
\includegraphics[width=0.33\textwidth]{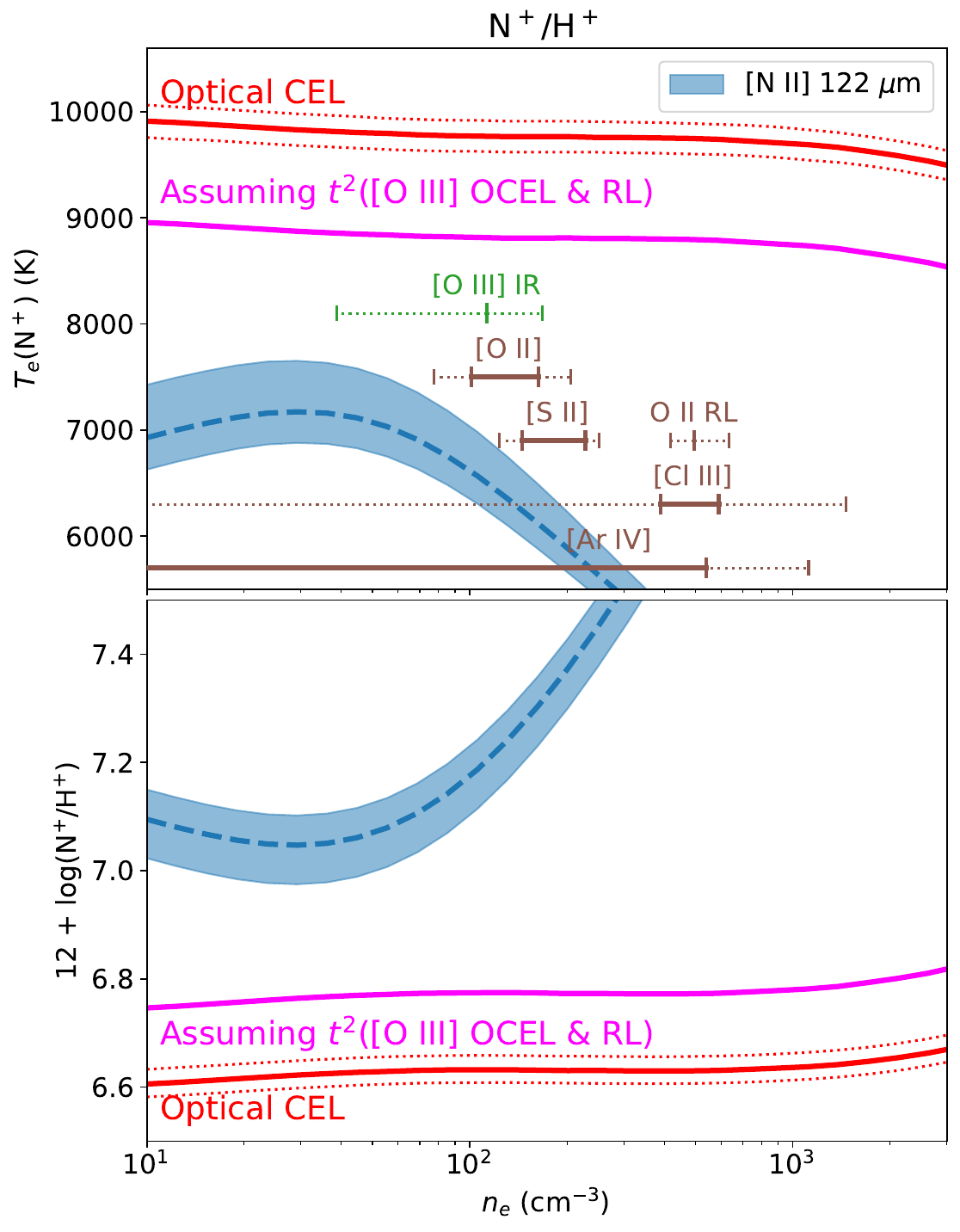}
\includegraphics[width=0.33\textwidth]{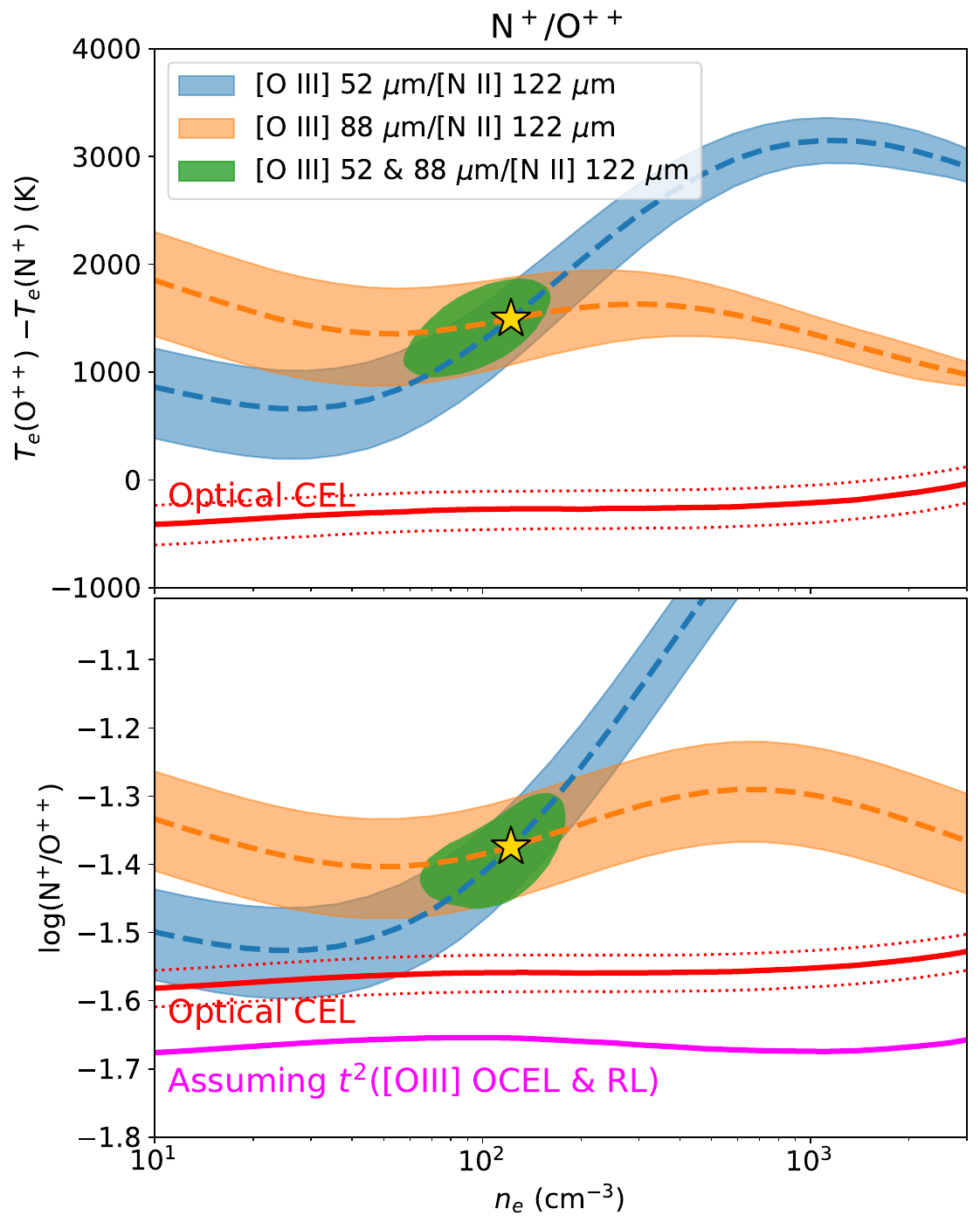}
\caption{  $T_e$ and abundance measurements of O$^{++}$/H$^+$ (\textit{left}), N$^{+}$/H$^+$ (\textit{middle}), and N$^+$/O$^{++}$ (\textit{right}) from various methods. The red solid and dashed lines represent the measurements and their 1$\sigma$ uncertainties obtained via the OCEL method. The magenta lines denote RL metallicity measurements for O$^{++}$/H$^+$ and the inferred ``true'' mean temperature ($T_0$) and metallicities for N$^+$/H$^+$ and N$^+$/O$^{++}$, assuming temperature fluctuations as deduced from the \Oiii\ RL vs. OCEL method. 
The effect of temperature fluctuations, parameterized by $t^2$, is such that the mean $T_0$ is lower than the $T_e$ derived from OCEL, while ion abundances O$^{++}$/H$^+$ and N$^{+}$/H$^+$ are higher. We expect the ratio N$^{+}$/O$^{++}$ to be lower than OCEL when accounting for temperature fluctuations (as shown in the lower right panel).
The measurements and 1$\sigma$ uncertainties derived from different far-IR lines are illustrated by the blue and orange lines and the corresponding shaded areas. The green shaded regions and yellow stars show the $1\sigma$ confidence interval and the best-fit value of $n_e$, derived from the combination of \Oiii\ $\lambda\lambda$52~\um\ and 88~\um\ lines. Additionally, $n_e$ measurements obtained from \Oii, the \Oiirec\ V1 RL multiplet, and \Sii\ are depicted with brown errorbars, in which the inner solid range indicates the aperture variation and the outer dotted ranges indicate the $1\sigma$ uncertainty. All density indicators suggest $n_e \lesssim 10^3$~cm$^{-3}$, corresponding to the regime in which \Oiii~$\lambda$52~\um\ and \Oiii~$\lambda$88~\um/\Nii~$\lambda$122~\um\ are relatively robust abundance indicators.
\label{fig:metallicities}}
\end{figure*}

Electron temperatures for O$^{++}$ and N$^+$ [$T_e$(O$^{++}$) and $T_e$(N$^+$)], along with their metallicities (i.e., $\mathrm{O}^{++}/\mathrm{H}^+$ and $\mathrm{N}^+/\mathrm{H}^+$ abundance ratios), were determined using \texttt{PyNeb} \citep{luridiana15}. We adopt collisional strengths from \citet{storey14, tayal11} and transition probabilities from \citet{storey00, fischer04}. Our approach is to measure $T_e$(IR) from the extinction-corrected ratios of far-IR to optical nebular lines (namely \Oiii\ $\lambda\lambda$52~\um/5007, \Oiii\ $\lambda\lambda$88~\um/5007, and \Nii\ $\lambda\lambda$122~\um/6583), and metallicities from the ratios of far-IR to \ion{H}{1} Balmer lines (e.g., \Oiii\ $\lambda$52~\um/H$\alpha$). We compute $T_e$(IR) and metallicities as a function of electron density $n_e$. We also measure $n_e$(IR) directly from the \Oiii\ $\lambda\lambda$52~\um/88~\um\ flux ratio. The 1$\sigma$ uncertainties are estimated using the Markov Chain Monte Carlo (MCMC) method, accounting for systematic uncertainties arising from flux calibration and extinction correction described earlier. The posterior distributions of $T_e$, $n_e$, and metallicities are illustrated in Fig.~\ref{fig:metallicities} with maximum-likelihood values reported in {Table \ref{tab:ne} for $n_e$ and Table \ref{tab:te_met} for $T_e$ and metallicities. We consider the flux calibration error (5\% for KCWI; 11\% for PACS and FIFI-LS), and the dust-correction errors in the posteriors.} In addition to the $T_e$(\Oiii) and $T_e$(\Nii) values used in subsequent derivations, we also report $T_e$(\Sii), $T_e$(\Oii), $T_e$(\Hei), as well as the $S^+$, $S^{++}$, and S abundances for reference. The $T_e$(\Sii) and $T_e$(\Siii) values are derived from the corresponding auroral-to-nebular line ratios. The $T_e$(\Hei) is measured from the \Hei~$\lambda\lambda 7281/6678$ ratio, with both lines originating from the singlet system, which is not susceptible to metastable states and is therefore considered reliable for temperature determinations \citep{md25}.

\begin{deluxetable*}{lccc}
\tablecaption{{Temperatures and metallicities of Haro~3} \label{tab:te_met}}
\tablehead{ Property & OCEL & Far-IR & RL}
\startdata
$T_e$(\Oiii) (K) & $9498_{-86}^{+95}$\tablenotemark{a} & $7960_{-260}^{+450}$ & ---  \\
$T_e$(\Nii) (K)& $9770_{-140}^{+150}$\tablenotemark{a} & $6540_{-430}^{+740}$\tablenotemark{a} & --- \\
$T_e$(\Oii) (K)& $11330_{-430}^{+410}$ & --- & --- \\
$T_e$(\Sii) (K) & $10480_{-560}^{+600}$ & --- & --- \\
$T_e$(\Siii) (K) & $8920_{-220}^{+250}$ & --- & --- \\
$T_e$(\Hei)\tablenotemark{b} (K) & $5650 \pm 530$ & --- & --- \\ 
$12 + \log(\mathrm{O}^{++}/\mathrm{H}^{+})$ & $8.202\pm 0.022$\tablenotemark{a} & $8.514_{-0.090}^{+0.051}$ & $8.440 \pm 0.022$\tablenotemark{a} \\
$12 + \log(\mathrm{O}^{+}/\mathrm{H}^{+})$ & {$7.648_{-0.072}^{+0.080}$} & --- & --- \\
$12 + \log(\mathrm{O}/\mathrm{H})$\tablenotemark{c} & $8.309 \pm 0.024$ & $8.569_{-0.072}^{+0.043}$ & $8.505 \pm 0.022$ \\
$12 + \log(\mathrm{N}^{+}/\mathrm{H}^{+})$& $6.632 \pm 0.024$\tablenotemark{a} & $7.14_{-0.11}^{+0.12}$\tablenotemark{a} & --- \\
$12 + \log(\mathrm{N}/\mathrm{H})$\tablenotemark{d} & $7.098 \pm 0.033$ & $7.60_{-0.11}^{+0.13}$ & ---\\
$12 + \log(\mathrm{S}^{++}/\mathrm{H}^{+})$ & $6.692_{-0.054}^{+0.050}$\tablenotemark{a} & --- & --- \\
$12 + \log(\mathrm{S}^{+}/\mathrm{H}^{+})$ & $5.839 \pm 0.071$\tablenotemark{a} & --- & --- \\
$12 + \log(\mathrm{S}/\mathrm{H})$\tablenotemark{e} & $6.828_{-0.054}^{+0.045}$ & --- & ---  \\
$\log(\mathrm{N}^{+}/\mathrm{O}^{++})$ & $-1.559 \pm 0.027$\tablenotemark{a} & $-1.372_{-0.091}^{+0.076}$ & --- \\
$\log(\mathrm{N}/\mathrm{O})$\tablenotemark{cd} & $-1.014 \pm 0.076$ & $-0.76_{-0.11}^{+0.10}$ & --- \\
$\log(\mathrm{S}/\mathrm{O})$ & $-1.481_{-0.059}^{+0.049}$ & --- & --- \\\hline
\enddata
\tablenotetext{a}{Assuming $n_e$(\Oiii\ IR). }
\tablenotetext{b}{Measured using the \Hei\ $\lambda\lambda 7281/6678$ ratio. }
\tablenotetext{c}{Assuming $\mathrm{O/H} = \mathrm{O}^{+}/\mathrm{H}^{+} + \mathrm{O}^{++}/\mathrm{H}^{+}$, where $\mathrm{O}^{+}/\mathrm{H}^{+}$ is measured from the OCEL. }
\tablenotetext{d}{Assuming $\mathrm{N/H} = \mathrm{N}^{+}/\mathrm{H}^{+} + \mathrm{N}^{++}/\mathrm{H}^{+}$, and $\mathrm{N}^{+} / \mathrm{N}^{++} = \mathrm{O}^{+} / \mathrm{O}^{++}$(OCEL). }
\tablenotetext{e}{Assuming $\mathrm{S/H} = \mathrm{ICF}(\mathrm{S}^{+} + \mathrm{S}^{++}) \times (\mathrm{S}^{+} + \mathrm{S}^{++})/\mathrm{H}^{+}$, where the ionization correction factor, ICF, is adopted from the theoretical relation with the $\mathrm{O}^{+}/\mathrm{O}$ ratio presented in \citet{dors16}}.
\end{deluxetable*}

\begin{deluxetable}{llcc}
\tablecaption{{The ADF and $t^2$ of Haro~3 \label{tab:t2}}}
\tablehead{ Ion & Quantity & Far-IR vs. OCEL & RL vs. OCEL }
\startdata
\multirow{3}{*}{O$^{++}$} & ADF (dex) &  $0.307^{+0.058}_{-0.087}$ & $0.238 \pm 0.029$ \\
& $T_0$ (K) & $7620^{+490}_{-370}$ & $8160 \pm 130$ \\
& $t^2$ & $0.078 \pm 0.027$ & $0.051\pm 0.010$ \\\hline
\multirow{3}{*}{N$^+$} & ADF (dex) & $0.52^{+0.13}_{-0.11}$ & --- \\
& $T_0$ (K) & $5400^{+1200}_{-1800}$ & --- \\
& $t^2$ & $0.54^{+1.11}_{-0.26}$ & --- \\
\enddata
\end{deluxetable}

The $T_e$ and metallicity values derived from far-IR emission lines are notably influenced by $n_e$ (Fig.~\ref{fig:metallicities}), due to the low critical density ($n_\mathrm{crit} \sim 10^2$--$10^3$ cm$^{-3}$) of these lines. The \Oiii\ $\lambda\lambda$52~\um/88~\um\ ratio provides a self-consistent measurement of $n_e$ within the same aperture and from the same \Oiii\ emitting gas. We adopt this $n_e$(\Oiii) for all $T_e$ and metallicity calculations in this work. For reference, in Table \ref{tab:te_met} and Fig. \ref{fig:metallicities}, we also include $n_e$ values measured from the optical \Oii\ $\lambda\lambda 3726$/$3729$ and \Sii\ $\lambda\lambda 6716$/$6731$ ratios, as well as those derived from the relative intensities of the \Oiirec\ V1 multiplet. For both \Oii\ and \Sii, $n_e$ has been determined from apertures with $r_\mathrm{ap} = 2\secpoint25$ and 3" before and after PSF-matching. The $n_e$ values measured from the smaller $r_\mathrm{ap} = 2\secpoint25$ aperture are higher than those from the 3" aperture, aligning with a density gradient in the \Hii\ region as reported in previous studies \citep{binette02, phillips07, garciabenito10, jin23}. We also provide constraints from the \Cliii\ and \Ariv\ doublets. However, due to their high $n_\mathrm{crit}$, the densities inferred from \Cliii\ and \Ariv\ are effectively upper limits, consistent with a range of $\simeq 0$--$1000~\mathrm{cm}^{-3}$. The difference between the densities measured from these two doublets and all other densities reported in Table \ref{tab:ne} are statistically insignificant. For example, the probabilities of $n_e$(\Oiii) being consistent with $n_e$(\Ariv) compared to $n_e$(\Oiirec\ RL) being consistent with $n_e$(\Ariv) are only different by $\simeq 20\%$. Assuming an $n_e = 1000~\mathrm{cm}^{-3}$, the $O^{++}/H^+$ metallicity derived from \Oiii\ $\lambda52~\um$ would only increase by 0.1~dex. All optical measurements [$n_e$(\Oii), $n_e$(\Sii), $n_e$(\Cliii), and $n_e$(\Ariv)] are consistent (within $<1\sigma$) with the far-IR $n_e$(\Oiii) measurements. However, even after accounting for aperture differences, the $n_e$ derived from the \Oiirec\ recombination lines remains considerably higher than the \Oii\ and \Sii\ measurements. 

The $T_e$ and metallicity are measured from OCELs and RLs in a manner similar to that used for the far-IR analysis. $T_e$(OCEL) is calculated from the auroral-to-nebular flux ratios (i.e., \Oiii\ $\lambda\lambda$4363/5007 and \Nii\ $\lambda\lambda$5755/6583). These temperatures are then used to derive CEL-based metallicities from the \Oiii\ $\lambda$5007/H$\alpha$ and \Nii\ $\lambda$6583/H$\alpha$ ratios. 
We also use $T_e$(O$^{++}$) obtained from OCELs to calculate the \Oiirec\ RL emissivity. The total flux from the \Oiirec\ lines at 4638.9, 4641.8, 4649.1, 4650.8, 4661.6, and 4676.2~\AA\ in the V1 multiplet is then used to calculate the RL-based O$^{++}$ metallicity. We note that RL emissivity and hence metallicity exhibits a relatively weak dependence on $T_e$. The $T_e$ and metallicities derived from OCELs and RLs have minimal sensitivity to electron density ($n_e$) in the low-density regime which applies here (Fig.~\ref{fig:metallicities}).

In addition to ion abundances, we also calculate the total O/H, N/H, and N/O ratios. The O/H ratio is derived by summing the O$^+$/H$^+$ and O$^{++}$/H$^+$ ratios, assuming that all oxygen in the \Hii\ regions is singly or doubly ionized. This is expected to be accurate to within a few percent \citep{izotov06, berg18}. The O$^+$/H$^+$ metallicity is measured from the direct-$T_e$ method using the \Oii\ $\lambda\lambda$3726,29, and the auroral \Oii\ $\lambda\lambda$7318,20,30,31 fluxes.
For N, we assume that the ratio of singly and doubly ionized N is the same as O, i.e., {N$^{++}$/N$^{+}$~$=$~O$^{++}$(OCEL)/O$^{+}$(OCEL) $=1.92$}.
The N$^{++}$/N$^+$ ratio implied from this assumption is in agreement with the \Niii\ $\lambda$57~\um/\Nii\ $\lambda$122~\um\ ratio reported in \citet{cormier15}. It is important to note that the total N and O metallicities presented in this study are intended only as reference values. Our conclusions regarding the ADF discussed in \S\ref{sec:discussion}, do not depend on these results.

Fig. \ref{fig:metallicities} summarizes the $T_e$ and ion abundance measurements as functions of $n_e$. We find a typical abundance discrepancy in Haro 3, with the RL-based abundance of O$^{++}$ higher than OCELs by $\sim 0.25$ dex. The far-IR \Oiii\ lines suggest a lower $T_e$ and higher abundance than OCELs, and are consistent within 1$\sigma$ uncertainty of the RL metallicity. This is in contrast with \citetalias{chen23}, in which we instead found the O/H metallicity derived from the far-IR and OCEL methods to be consistent. 
{To infer temperature fluctuations, we adopt the formalism of Equation \ref{eq:temperature_fluctuations} developed by \citet{peimbert67}, assuming a Gaussian distribution of $T_e$. Following \citetalias{chen23}, $t^2$ is constrained via MCMC fitting to dust corrected \Oiii, \Oiirec\ RL, or \Nii\ fluxes using $T_0$, $t^2$, and an arbitrary flux normalization factor as free parameters. In the regime of small $t^2$ ($t^2 \ll 1$), the formalism reduces to Equation 18 of \citet{peimbert67}. }
Our results for Haro 3 can be explained by temperature fluctuations and suggest $t^2 \sim 0.05$ (Table~\ref{tab:t2}). 
However, the far-IR \Nii\ emission does \textit{not} support a simple temperature fluctuation scenario. A much higher $t^2 \sim 0.5$ and a much lower average temperature $T_0$ are required to reconcile the optical and far-IR CELs of \Nii. {In addition to this unrealistically large $t^2$, the \Nii\ measurement differs from the two \Oiii\ measurements by $\sim 4\sigma$ in the 2D space of $(t^2, T_0)$ (see \S\ref{sec:discussion_t2} and Fig. \ref{fig:t2_T0}).} {Note that for $t^2$ comparable to 1, the $T_e$ distribution has to be truncated at $T_e = 500~\mathrm{K}$, a lower bound imposed by \textsc{PyNeb}. 
The N$^+$/O$^{++}$ abundance ratio offers another test. The far-IR N$^+$/O$^{++}$ ratio is $\simeq 2\sigma$ higher than that derived from OCELs. However, since the \Nii\ $\lambda$6583 emission is less sensitive to $T_e$ than \Oiii\ $\lambda$5007, a similar $t^2$ for N$^+$ compared to O$^{++}$ should result in a ``true'' N$^+$/O$^{++}$ ratio that is lower than the value we report from the OCELs; the far-IR value is likewise expected to be lower. Instead, our far-IR N$^+$/O$^{++}$ ratio is $\simeq 3\sigma$ higher than expected under the assumption that $t^2(\mathrm{N}^+) = t^2(\mathrm{O}^{++})$ (Fig.~\ref{fig:metallicities}, lower right panel). Note that the 2\secpoint25 and 3" apertures have a difference of $\simeq 0.05$ dex in the $O^{++}/H^+$ ratios. However, this difference is not sufficient to explain the difference in the optical and IR $N^+/O^{++}$ measurements, and it would further increase the discrepancy for the $O^{++} / H^+$ abundance.} 
While the IR-to-optical comparisons are subject to various sources of uncertainty (e.g., attenuation correction, electron density, flux calibration), the purely IR measurement of N$^+$/O$^{++}$ is relatively robust. However, despite the fact that the far-IR flux calibration uncertainties are relatively robust (see \S\ref{sec:flux_measurements}), if the flux calibration uncertainty increases from the fiducial 11\% to 20\%, the inconsistency between the far-IR and optical N$^+$/O$^{++}$ measurements would be reduced to $\sim 1.5\sigma$. However, the inconsistency between the \Nii\ and \Oiii\ $(t^2, T_0)$ measurements would remain at a high significance of $>3\sigma$. 

\section{Discussion\label{sec:discussion}}

\subsection{To $t^2$ or not to $t^2$?\label{sec:discussion_t2}}

\begin{figure*}[ht!]
\centering
\includegraphics[width=0.9\textwidth]{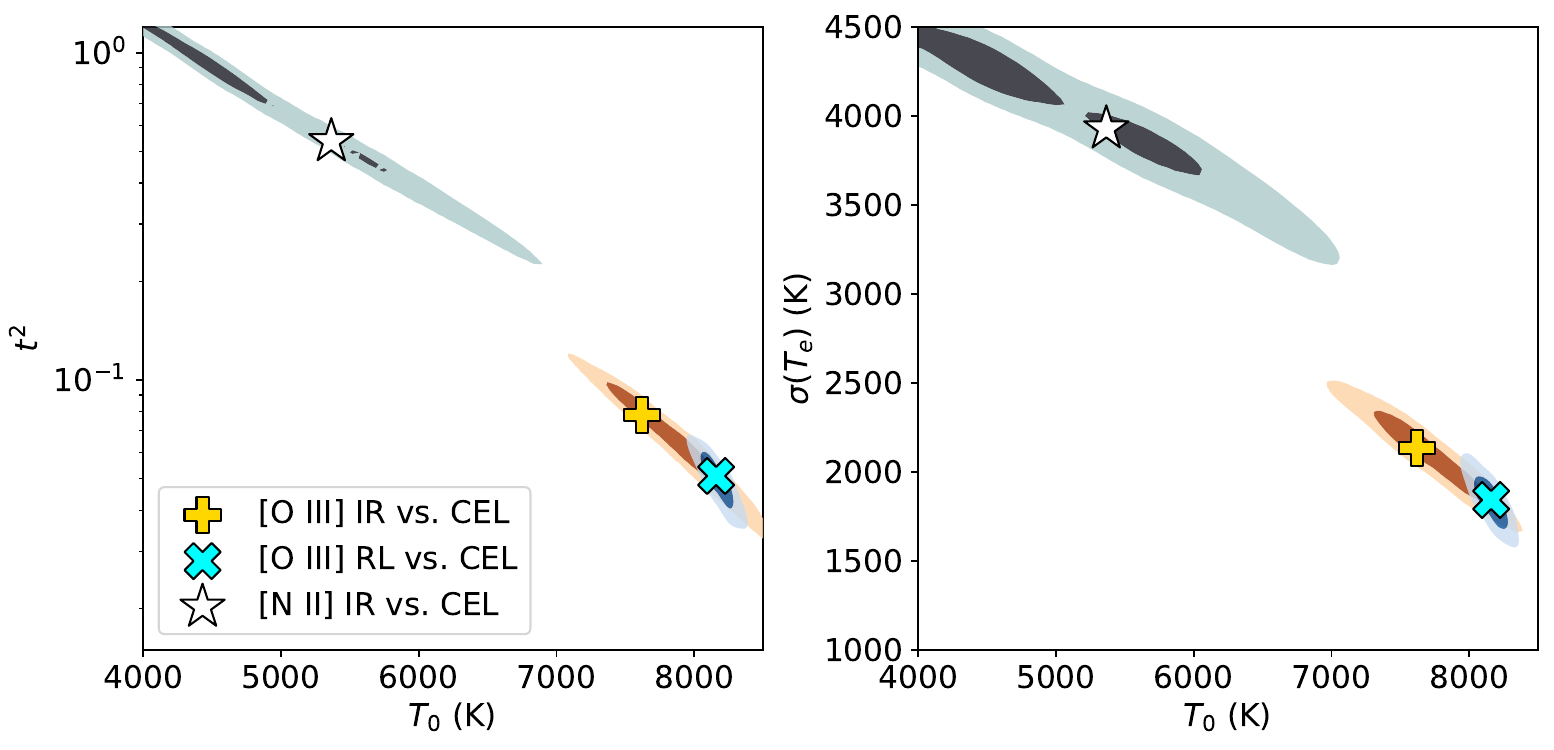}
\caption{  The posterior distributions of temperature fluctuations (expressed as $t^2$, \textit{left}; or $\sigma(T_e)$, \textit{right}) and the average temperatures ($T_0$), as deduced from the ADFs between various methods under the assumption that ADFs are solely caused by temperature fluctuations. Symbols show the best-fit values, while shaded regions denote the $1\sigma$ and $2\sigma$ confidence intervals. The two methods based on \Oiii\ emission demonstrate consistency within a $1\sigma$ range. However, in the case of \Nii, the inferred values are significantly inconsistent {($\sim 4\sigma$)} compared to the those obtained from \Oiii, and the implied temperature fluctuations are considerably larger, with $\sigma(T_e)$ comparable to $T_0$.
\label{fig:t2_T0}}
\end{figure*}

A key objective of this study is to assess the degree to which temperature fluctuations can account for the ADF, which is ubiquitously observed between OCEL and RL. 
We find agreement in O$^{++}$/H$^+$ abundance between the RL and far-IR measurements, which appears to support the hypothesis that the ADF is caused by temperature fluctuations. 
However, the N$^+$/H$^+$ and N$^+$/O$^{++}$ ratios derived from far-IR and OCEL observations \textit{do not support} this {relatively simple interpretation of the temperature fluctuations paradigm (Figs.~\ref{fig:metallicities} \& \ref{fig:t2_T0}). }

With the goal to understand whether temperature fluctuations alone can explain the disparity, we convert the differences in metallicity measurements between RL vs. OCEL and far-IR vs. OCEL into pairs of $(t^2, T_0)$ based on Eq.~\ref{eq:temperature_fluctuations}. Traditionally, such calculations have been performed analytically, under the assumption that temperature fluctuations are relatively minor compared to $T_0$ (see \citealt{peimbert17}). However, this assumption proves inadequate for reconciling the \Nii\ far-IR and  OCEL measurements. Consequently, we performed a numerical integration of Eq. \ref{eq:temperature_fluctuations} and estimated uncertainties in the $(t^2, T_0)$ parameters using MCMC.

In Fig. \ref{fig:t2_T0} and Table \ref{tab:te_met}, we present the posterior distributions of ($t^2$, $T_0$) calculated from comparisons of \Oiii\ RL vs. OCEL, \Oiii\ far-IR vs. OCEL, and \Nii\ far-IR vs. OCEL. As expected from the the \Oiii\ RL and far-IR metallicity measurements, the $(t^2, T_0)$ values derived from \Oiii\ analyses are in mutual agreement. The $t^2$ value obtained from \Oiii\ RL vs. optical CEL aligns with the typical range of $t^2$ ($\simeq 0.02$--0.2) reported in similar \Hii\ regions \citep{peimbert05, esteban09, esteban14}. Additionally, our findings are consistent with typical \Hii\ regions in [$t^2$, $T_e$(\Oiii)$-T_e$(\Nii), $\log(\mathrm{O/H})$] space, which has been presented as evidence for temperature fluctuations \citep{md23}. The $t^2$ value derived from far-IR vs. OCEL is consistent within $\sim 1\sigma$ of the RL vs. OCEL value. 

However, the \Nii\ far-IR vs. OCEL comparison presents a more complex picture. The derived $T_0$ is consistent with the \Hei\ temperature measured from the \Hei\ $\lambda\lambda 7281/6678$ ratio, supportive of the temperature fluctuation scenario, but the ($T_0$, $t^2$) values differ by {$\sim 4\sigma$} from the \Oiii\ values, with an anomalously large $t^2$ (Table~\ref{tab:t2}). Some degree of difference is expected since N$^+$ and O$^{++}$ originate from fundamentally different regions within the \Hii\ regions, consistent with their distinct ionization energies (O$^{++}$: 35.12 eV; N$^+$: 14.53 eV) and corroborated by spatially resolved photoionization models \citep[e.g., ][]{sankrit00, arthur11, yang23, jin23}. However, the $t^2$ from \Nii\ far-IR vs. OCEL corresponds to a temperature standard deviation $\sigma(T_e) = 3930_{-490}^{+670}~\mathrm{K}$, or approximately 65--85\% of $T_0$. \citet{stasinska00} noted that traditional photoionization models struggle to account for the degree of temperature fluctuations indicated by the ADF, suggesting that additional factors like cosmic rays \citep{giammanco05} may be needed to explain fluctuations on the order of even $\simeq 1000$--2000~K. {Given the extent of temperature fluctuations implied by our measurements of both \Oiii\ and \Nii\ in Haro 3, it imposes a substantial constraint on the source of temperature fluctuations as an explanation of the ADF.}

\subsection{Non-thermal effects}

Another hypothesis proposed to explain the ADF involves the deviation from thermal equilibrium of electrons, leading to a distribution of electron energies that diverges from the classical Maxwell-Boltzmann framework. \citet{nicholls12, nicholls13, dopita13} investigated the concept of a $\kappa$-distribution, a generalized Lorentzian model characterized by an extended high-energy tail, as a potential mechanism underlying the ADF observed in \Hii\ regions and planetary nebulae. The $\kappa$-distribution has been documented within the solar system, primarily attributed to the influx of high-energy plasma from the solar wind. This mechanism suggests that a higher proportion of high-energy electrons could enhance the population of ions in excited states, leading to an overestimation of $T_e$ when derived from the auroral to nebular line ratios through the ``direct-$T_e$'' method. {However, \citet{ferland16} pointed out that theoretically, a non-thermal distribution of electrons cannot exist in photoionized nebulae, since the thermalizing timescale is $\sim 10^{-8}$ times the heating timescale. }

We can test whether this offers a satisfactory explanation thanks to the difference in $T_e$-sensitivity of the \Oiii\ $\lambda\lambda$4363/5007 and \Nii\ $\lambda\lambda$5755/6583 ratios, which likewise respond differently to the $\kappa$-distribution. Specifically, to explain the ADF, the $T_e$(\Oiii) of Haro~3 requires a less pronounced $\kappa$-distribution ($\kappa \gtrsim 50$)\footnote{The parameter $\kappa$ is defined such that a lower value of $\kappa$ has a more pronouced high-energy tail, while $\kappa = \infty$ is equivalent to the Maxwell-Boltzmann distribution.} {that would be indinstinguishable from temperature fluctuations effects \citep{peimbert13}}, while $T_e$(\Nii) suggests a significant modulation from the high-energy tail ($\kappa \sim 10$). Since N$^+$ and O$^{++}$ ions are not strictly co-spatial, it is, again, reasonable to expect different $\kappa$-distributions derived from the N$^+$ and O$^{++}$ gas. However, the implication that N$^+$ requires more high-energy injection (based on its lower associated $\kappa$ value) is contradictory to the naive picture that gas closer to the ionizing energy source is more highly ionized. Therefore, we conclude that a non-thermal $\kappa$-distribution of electron energies is unlikely to be the cause of the ADF.

\subsection{Density imhomogeneity}

{In addition to temperature fluctuations, \citet{md23ne} and \citet{rv23} have recently discussed the effects of density ($n_e$) fluctuations in \Hii\ regions. Variations in $n_e$ are particularly relevant to this work since the far-IR lines have small critical densities ($n_\mathrm{crit} \sim$ a few hundred $\mathrm{cm}^{-3}$). Such $n_\mathrm{crit}$ would raise concerns about whether the inclusion of high-density clumps or the diffuse ionized gas (DIG) from the host galaxies, in addition to temperature fluctuations, could significantly offset the metallicity measurements. In this section, we examine the extent to which density inhomogeneity can plausibly explain the apparent discrepancies. }

One potential effect of density inhomogeneity is that high density gas is essentially hidden from the CEL emission with relatively low $n_\mathrm{crit}$. For example, the $n_e$ values derived from \Oii\ and \Sii\ nebular emission are reported to be approximately 1--2 dex lower than those obtained from auroral or UV lines (\citealt{mingozzi22}). In the case of the \Oiii\ 52~\um\ and 88~\um\ lines, their $n_\mathrm{crit}$ are $\simeq 3500$ and $500~\mathrm{cm}^{-3}$ respectively. Therefore, any gas with $n_e \gtrsim 3500~\mathrm{cm}^{-3}$ would be hidden from those lines, but is still visible by the \Oiii\ OCEL and the H Balmer emission. This effect has recently been observed in the $z>6$ galaxies \citep{usui25, harikane25}. However, we note that if such effect exists in Haro~3, it would increase the discrepancy in the mesured $O^{++}/H^+$ metallicity between the optical and the far-IR lines, as the true metallicity would be even higher than the far-IR measured metallicity.  We can also examine the effect by the $n_e$ constraints from the optical diagnostics with higher critical densities (\Cliii: $\simeq 0.7$--$3.9\times 10^4~\mathrm{cm}^{-3}$ and \Ariv: $\simeq 1.5\times 10^4$--$1.3\times 10^5~\mathrm{cm}^{-3}$). These diagnostics yields an upper limit of $\sim 1000~\mathrm{cm^{-3}}$, confirming that the majority of the gas in this system has $n_e < 1000~\mathrm{cm^{-3}}$, consistent with the \Oii, \Sii, and \Oiii\ $n_e$ measurements.

The accuracy of our metallicity measurements, especially for the far-IR emission, may also include contributions from the diffused ionized gas (DIG) of the host galaxies. At low $n_e$, DIG would be particularly visible for the far-IR emission due to their low $n_\mathrm{crit}$. However, this effect should be minor. The majority of the \Hii\ gas has a low $n_e \lesssim 10^3~\mathrm{cm}^{-3}$ based on the \Cliii\ and \Ariv\ upper limits. At this range, DIG ad HII regions are equally indistinguishable to both the far-IR and OCEL emission lines. Additionally, the H$\alpha$ surface luminosity for the PSF-degraded 3" aperture is $\log (\Sigma_{H\alpha} / \mathrm{erg~s}^{-1}~\mathrm{kpc}^{-1}) \simeq 40.8$, and the equivalent width is $\mathrm{EW}(H_\alpha) \simeq 320~\mathrm{\AA}$. Both suggest that the nebular emission is strongly dominated by \Hii\ regions \citep{zhang17, sanders17}. By comparing the H$\alpha$ fluxes in the extraction aperture and a region on the opposite side of the host galaxy, the contribution of DIG is estimated to be $< 1\%$ of the line fluxes. These comparisons suggest that the \Hii\ region dominates the host galaxy in nebular and ionizing radiation within both apertures. Therefore, density variations alone cannot be the major contributor for the inconsistent metallicity measurements.  

\begin{figure}[ht!]
\centering
\includegraphics[width=0.95\columnwidth]{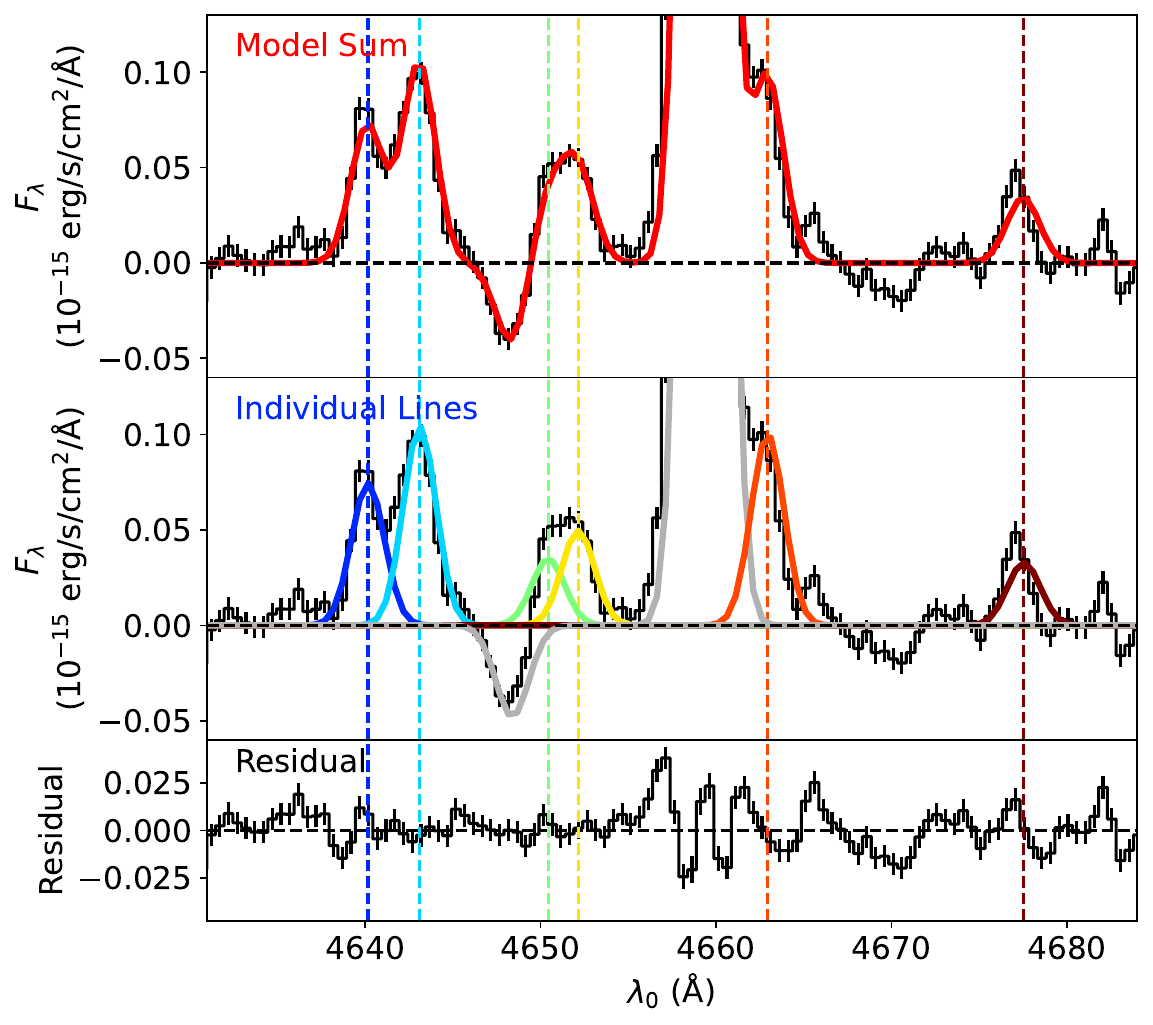}
\caption{   Optimal model fit from MCMC analysis for the \Oiirec\ V1 RL multiplet. The top panel presents the observed continuum-subtracted spectra in black, and the model sum that consists of multiple Gaussian components in red. The middle panel illustrates the decomposition of the total model into its constituent Gaussian components: those associated with the \Oiirec\ multiplet are highlighted in color, while those corresponding to other spectral features are shown in gray. The unidenfied absorption feature at rest-frame vacuum wavelength of $\lambda_0 = 4648.7~\mathrm{\AA}$ also appeared in \citetalias{chen23} as an emission line. We suspect that it may correspond to a diffuse interstellar band (DIB). The bottom panel displays the residuals of the fit. The residual is low (standard deviation $\times$ pixel size is $ <1\%$ of the total \Oiirec\ RL flux), indicating a good fit. 
\label{fig:o2_rl_fit}}
\end{figure}

\begin{figure}[ht!]
\centering
\includegraphics[width=0.95\columnwidth]{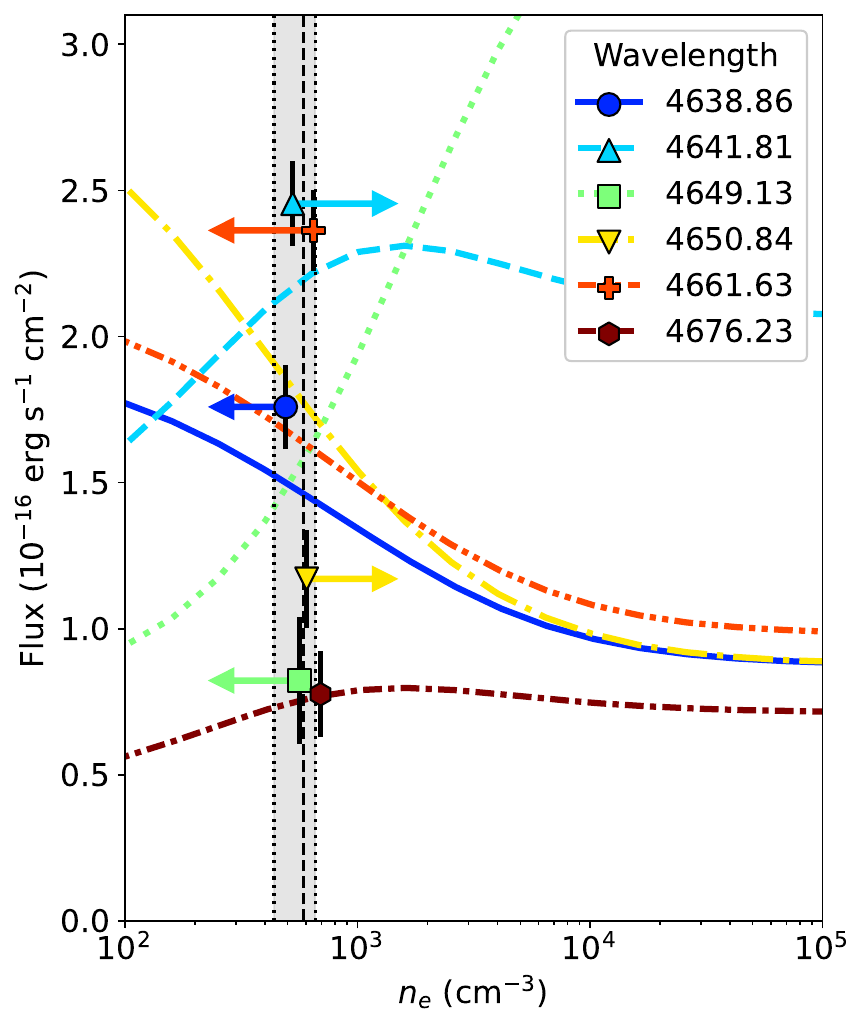}
\caption{ Observed fluxes and their $1\sigma$ uncertainties for the \Oiirec\ RL V1 multiplet (scattered points with error bars) are compared to theoretical predictions of line fluxes as a function of $n_e$ (curves). The colors match the individual components in Fig. \ref{fig:o2_rl_fit}. The best-fit $n_e$ value and its $1\sigma$ uncertainty, as determined by MCMC, are highlighted by vertical lines and a gray shaded area. To enhance clarity, the scattered points are deliberately displaced from the best-fit $n_e$ along the x-axis by increments of 0.03 dex. Arrows pointing from the measured fluxes indicate the necessary adjustments in $n_e$ to reconcile observed values with theoretical expectations. The theoretical curves are based on \citet{storey17} and are normalized to ensure the total flux at each $n_e$ aligns with the observed flux, highlighting potential contamination that may compromise the accuracy of the measured \Oiirec\ fluxes and, consequently, the inferred metallicity.
\label{fig:o2_rl_ne}}
\end{figure}

Within the framework of density fluctuations, an observed underestimation of $n_e$ by $\sim 300~\mathrm{cm}^{-3}$ aligns with estimates derived from \Oiirec\ RLs, as reported in Table \ref{tab:ne}. However, we find that the recombination coefficients for \Oiirec\ from \citet{storey17} cannot fit the observed fluxes of individual lines well (Appendix \ref{sec:line_fluxes}). The $n_e$(\Oiirec) value provided in Table \ref{tab:ne} reflects the most likely solution from MCMC based on current atomic data, yet this best-fit scenario (Fig. \ref{fig:o2_rl_fit}) overestimates the fluxes of \Oiirec\ 4649.13 and 4650.84 by $\sim 40\%$ (Fig. \ref{fig:o2_rl_ne}). This finding contrasts with the studies by \citetalias{chen23} on Mrk~71 and \citet{storey17} on various nearby \Hii\ regions, where adjusting $n_e$ yielded reasonable fits to the \Oiirec\ RL line ratios. As shown in Fig.~\ref{fig:o2_rl_fit}, several \Oiirec\ RLs are blended with the [\ion{Fe}{3}] $\lambda 4658$ feature and an unidentified absorption feature at $4648.7$~\AA. We suspect that this absorption may arise from diffuse interstellar bands (DIBs); however, its wavelength does not coincide with any entries in existing DIB catalogues, except for a candidate at 4650.77~\AA that could overlap with the \Oiirec\ $\lambda 4650.84$ line \citep{fan19}. This candidate was identified along significantly more dust-attenuated sightlines with a detection rate of $<20\%$, and the feature would need to be at least an order of magnitude broader to produce noticeable absorption at $\sim 4648.7$~\AA. Additionally, we do not detect any nearby strong DIB absorptions (e.g., at 4429.3~\AA\ or 4727.0~\AA) in our spectra. Nevertheless, this suggests that the \Oiirec\ $\lambda\lambda 4649.13$ and 4650.85 features may be contaminated by unidentified sources. We visually inspected model-generated spectra at $n_e \simeq 240$--3500~$\mathrm{cm}^{-3}$ and found that the modeled 4649.13 or 4650.84 fluxes are systematically overpredicted by at least a factor of $\sim 2$ relative to the observations. Therefore, we reconducted the same analysis with only a subset of the \Oiirec\ emission of 4638.86, 4641.81, and 4676.23 which are relatively robust. This subset suggests an $n_e = 440^{+230}_{-130}~\mathrm{cm}^{-3}$ and increases the ADF between \Oiii\ OCEL and RL by $\simeq 0.1$ dex. While the O/H metallicity from this subset is closer to the far-IR metallicity, the new $n_e$ is still inconsistent with {\Oii\ and \Sii\ } $n_e$ measurements. In another attempt to reconcile this discrepancy, we used a two-phase model (similar to the ones in \S\ref{sec:analytic_models} below) to explain the \Oiirec\ RL flux ratios. Nonetheless, all models generally favored minimal $n_e$ fluctuations and were unable to achieve a more accurate fit. These results do not necessarily imply issues with the O$^+$ atomic data. Instead, they more likely indicate that either the potential DIB contamination is substantially stronger than assumed or that a more complex density structure is required. In either case, this discrepancy underscores a fundamental challenge in using \Oiirec\ recombination-line fluxes for metallicity measurements and suggests that \Oiirec\ RLs alone do not provide robust constraints on density fluctuations.

The observed differences in $n_e$ across two extraction apertures for Haro 3, as documented in Table \ref{tab:ne}, indicate the presence of a moderate $n_e$ gradient ($\sim 70~\mathrm{cm}^{-3}$, or $\simeq 0.2$ dex in $\simeq 1"$). {However, this effect is negligible compared to the ADF. If we consider log-normal density distributions, such a standard deviation of $\simeq 0.3$ dex can lead to an overestimation of $T_e$ by $\sim 100$ K and an underestimation of the O$^{++}$/H$^+$ ratio by $\sim 0.02$ dex for the far-IR lines. The impact is also in the \textit{opposite direction} of the observed discrepancy between far-IR and optical measurements.} 

{In summary, we find that $n_e$ flutuations, which exist at a modest level, are not the dominant cause of the observed abundance discrepancies.}

\subsection{{Two-phase models}}
\label{sec:analytic_models}

In previous sections, we demonstrated that pure temperature and density fluctuations cannot explain the observed discrepancy between the OCEL, RL, and far-IR line flux ratios for \Oiii\ and \Nii. Meanwhile, inclusions of high-density clumps have been proposed as an alternative cause of the ADF \citep{stasinska07}, especially in planetary nebulae \citep[e.g., ][]{tsamis04, liu06, wesson08}. However, the extreme chemical inhomogeneities observed in planetary nebulae arise from unmixed stellar ejecta, and such conditions are not expected in \Hii\ regions. We therefore first consider restricted models in which both components share identical abundances but differ in their thermodynamic properties, while exploring whether bimodal distributions associated with two distinct phases can account for the observed properties of Haro~3.
The effect of bimodality is two-fold: 1) the bimodal distribution of the physical properties like $T_e$ can generate stronger observational effects compared to the Gaussian distribution assumed from the fluctuations models; and 2) the correlations in physical properties such as $T_e, n_e$, metallicity, and dust attenuation can modulate the observed line fluxes in ways that are difficult to reproduce with smooth fluctuation distributions.
In this section we describe a series of analytic models with two phases having distinct physical properties. We compare predicted fluxes from the sum of the two phases with those observed in Haro 3, specifically considering the \Oiii\ OCEL, the total \Oiirec\ RL, and the two far-IR \Oiii\ emission line fluxes. The \Nii\ $\lambda 6583$ / \Oiii\ $\lambda 5007$, and \Nii\ $\lambda 88~\um$ / \Oiii\ $\lambda 122~\um$ ratios are also calculated from the models as references, but have less constraining power for properties such as $n_e$ and are not fitted to the observed values (see details below).

We first focus on the $T_e$ and $n_e$, following the $T_e$ and $n_e$ fluctuations scenarios.  Given the limited constraints available from the observed line fluxes, we restrict the set of free parameters according to physically motivated considerations. \textit{Model A} assumes that the two components are in thermal pressure equilibrium, i.e., the electron temperatures and densities of the two components follow $T_1 n_1 = T_2 n_2$.  The assumption of thermal pressure equilibrium is often invoked when modeling stratified or multiphase gas in \Hii\ regions in small scales \citep{peguipnot01, pellegrini11}, and large-scale isobaric equilibrium are often assumed in photoionization models \citep{ferland17, kewley19}. However, we note that the pressure balance in small scale is not well-understood. Model A only represents one variation of our assumptions. We also assume that both components have the same dust attenuation (\chb) and O$^{++}$/H$^+$ metallicity. The fractional contributions of the two components ($f_1$ and $f_2$) are defined as the relative $O^{++}$ ion abundances contributed by each component. Since there are only two distinctive phases, $f_1 + f_2 = 1$. The minimum temperatures are set at 500~K, and the minimum densities at $1~\mathrm{cm}^{-3}$, both required by \textsc{PyNeb}.  An additional flux normalization factor $F_{H\beta}$ is introduced to normalize the model fluxes in order to fit the observed line fluxes. This results in 7 parameters, $[T_1, n_1, n_2, 12+\log(\mathrm{O}^{++}/\mathrm{H}^+), \chb, f_2, F_{H\beta}]$. However, the value of \chb\ is constrained by a prior set by the \chb\ value measured from the H Balmer and Paschen flux ratios (\S\ref{sec:attenuation}). Thus, the model has 6 free parameters, fitting 6 line fluxes (\Oiii\ $\lambda\lambda 4363, 5007, 52~\um, 88~\um$, \Oiirec\ RL, and H$\beta$). For the \Nii/\Oiii\ line ratios provided for reference, we assumed a constant $Te$(\Nii) derived from the OCEL \Nii\ $\lambda\lambda 5755/6583$ ratio,  {and an $N^+/O^{++}$ ratio that equalizes the observed and model predicted \Nii\ $\lambda6583$/\Oiii\ $\lambda5007$ ratio}. 
We do not attempt to fit the \Nii/\Oiii\ flux ratios because a reaonsable two-phase model would require more free parameters than the available observational constraints (i.e., \Nii\ line fluxes).
For example, the $N^+/O^{++}$ ratio likely changes in the two components due to the difference in ionization levels and enrichment history, and the emperical $T_e$(\textrm{\Nii})-$T_e$(\textrm{\Oiii}) may not work for the individual components. This analysis highlights the need to perform similar analyses with additional emission features such as those in the mid-IR, whose abundances are less sensitive to the enrichment history and ionization levels. 

\begin{figure*}[ht!]
\centering
\includegraphics[width=0.95\textwidth]{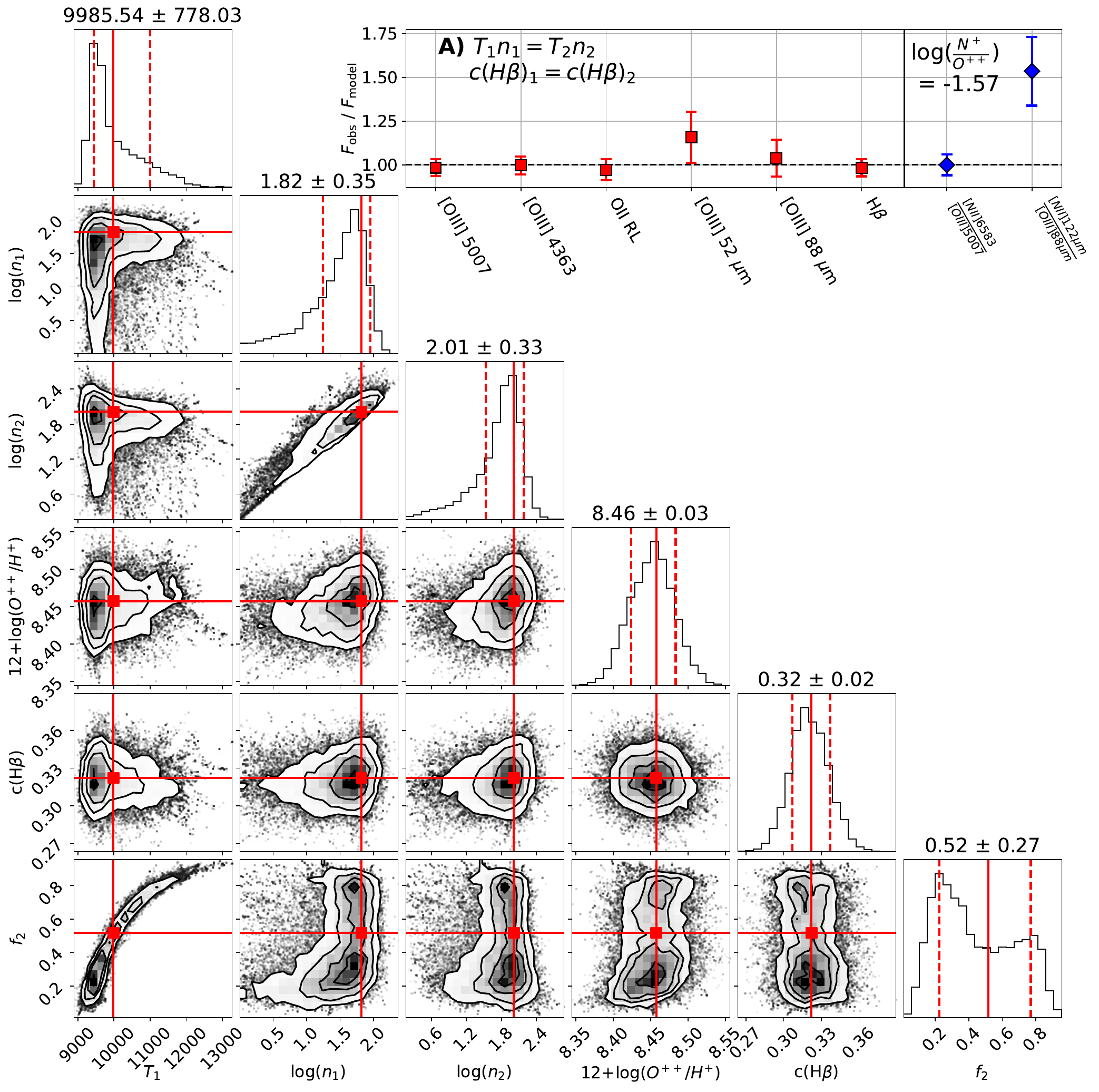}
\caption{ {Fitting results of the two-phase Model A, which assumes $T_1 n_1 = T_2 n_2$ and $\chb_1 = \chb_2$. Panels on the bottom-left show the correlations in the posterior distribution of the model parameters. The red solid lines indicate the best-fit values. The best-fit values and 1-$\sigma$ uncertainties are shown as the solid and dashed lines, with values given above the 1D histograms. The ratios between the best-fit vs. observed line fluxes are shown in the top-right panel. Fluxes shown as red square points are used to fit the model. The blue diamond points are instead provided for reference and verification, but are not used to fit the model. The posterior distribution of the flux normalizing factor $F_{H\beta}$ is omitted from this figure and has no correlation with other fitted quantities. }
\label{fig:corner_a}}
\end{figure*}

Fig.~\ref{fig:corner_a} summarizes the fitting results for Model A. The model prefers a consistent $n_e$ between the two phases, with a difference too small to cross $n_\mathrm{crit}$ of the far-IR emission lines and cause significant impact on the observed line ratios.  Meanwhile, the best-fit $T_2 = 6400 \pm 1800~\mathrm{K}$, $>2\sigma$ lower than $T_1$ ($1,0000 \pm 800$~K), indicating that in this system, $T_e$ variation is the driving explanation of the ADF. The amount of difference between $T_1$ and $T_2$ is $\sim 40\%$ of the mean $T_e$, about twice the level of $t^2$ required in the temperature fluctuations scenario. The effect on the line ratios is similar to the pure temperature fluctuations scenario, in that the OCEL fluxes are biased by the high-$T_e$ phase, and the $O^{++}/H^+$ ratios measured from the \Oiirec\ RL and the far-IR lines are accurate. However, the model underpredicts the \Oiii\ $\lambda\lambda 52~\um / 88 ~\um$ ratio by $\sim 15\%$ (or $\sim 1\sigma$), and the observed \Nii\ $\lambda~122~\um$ / \Oiii\ $\lambda~88~\um$ ratio is $\sim 50\%$ higher than the model prediction. This result is also consistent with the temperature fluctuations scenario, where we found that the far-IR \Nii\ $\lambda 122~\um$ is too high to be explained by temperature inhomogeneities. 

\begin{figure*}[ht!]
\centering
\includegraphics[width=0.95\textwidth]{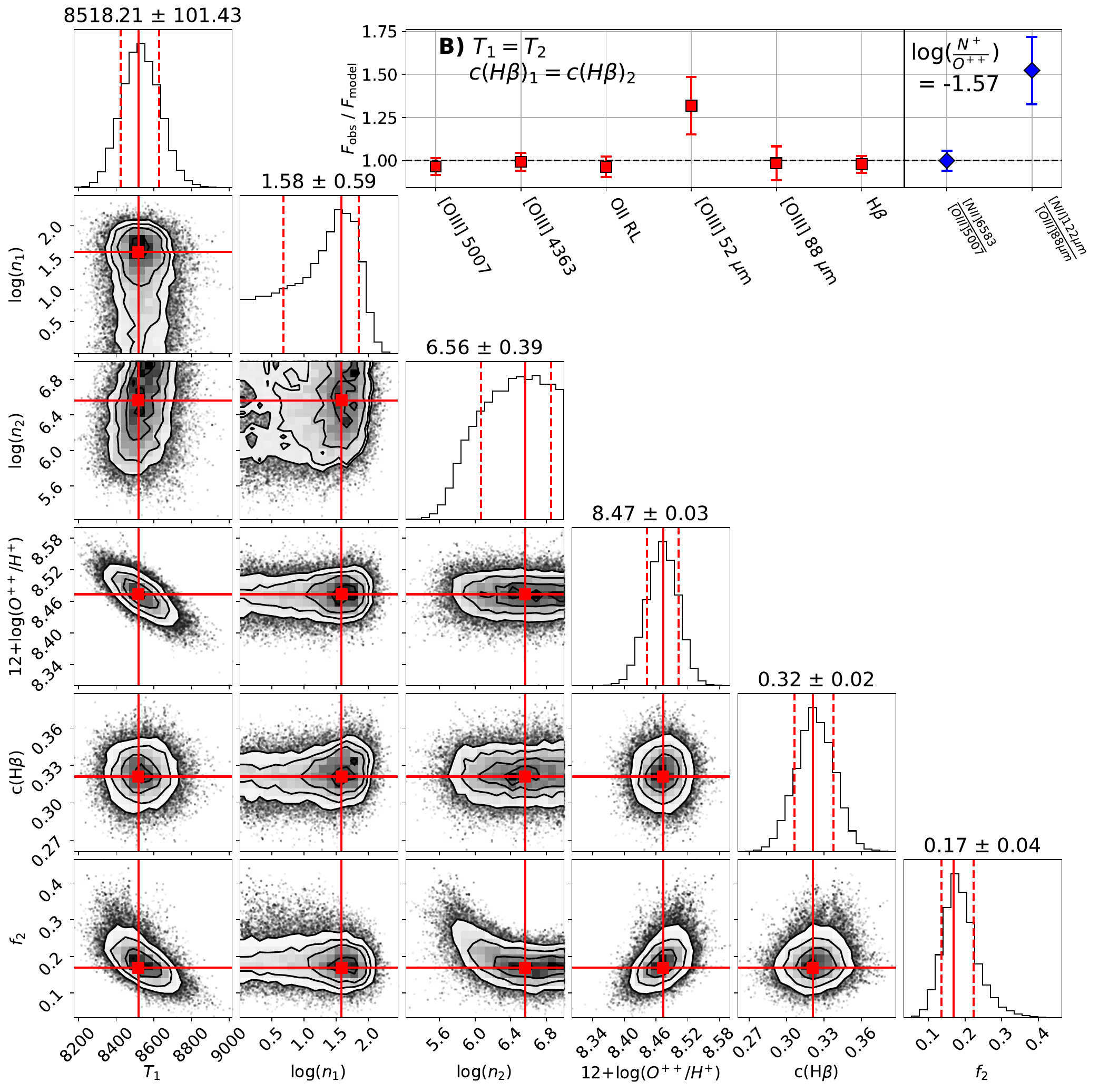}
\caption{ { Same as Fig. \ref{fig:corner_a}, but for Model B which assumes $T_1 = T_2$ and $\chb_1 = \chb_2$.}
\label{fig:corner_b}}
\end{figure*}

To check the effect of density bimodality, we set up \textit{Model B}, which assumes that the two phases share the same temperature ($T_1 = T_2$), while the densities are distinctive. The best-fit model  {confirms the expectation that,} to significantly modulate the measured line fluxes in OCEL and far-IR, the two components need to have $n_e$ different enough to be across the $n_\mathrm{crit}$. The best-fit results (Fig. \ref{fig:corner_b}) suggest a very dense ($10^{6.56} \mathrm{cm}^{-3}$) component consisting of $\simeq 17\%$ of the total $O^{++}$ gas. Similar to Model A, this model provides a reasonable (difference $<1\sigma$) fit to most fluxes, except for \Oiii~$\lambda52~\um$, for which the model underpredicts the flux by $\sim 30\%$ ($\sim 2\sigma$). This discrepancy results in the $n_e$ measured from the model-predicted \Oiii\ $\lambda\lambda 52~\um/88~\um$ ratio to be $50~\mathrm{cm}^{-3}$, about $1\sigma$ lower than the current measurement, and significantly ($>2\sigma$) lower than $n_e$(\Oii) and $n_e$(\Sii). This model also underpredicts the \Nii\ $\lambda 122~\um$ / \Oiii\ $\lambda 88~\um$ ratio similarly to Model A, and is overall a worse fit compared to Model A.  

\begin{figure*}[ht!]
\centering
\includegraphics[width=0.95\textwidth]{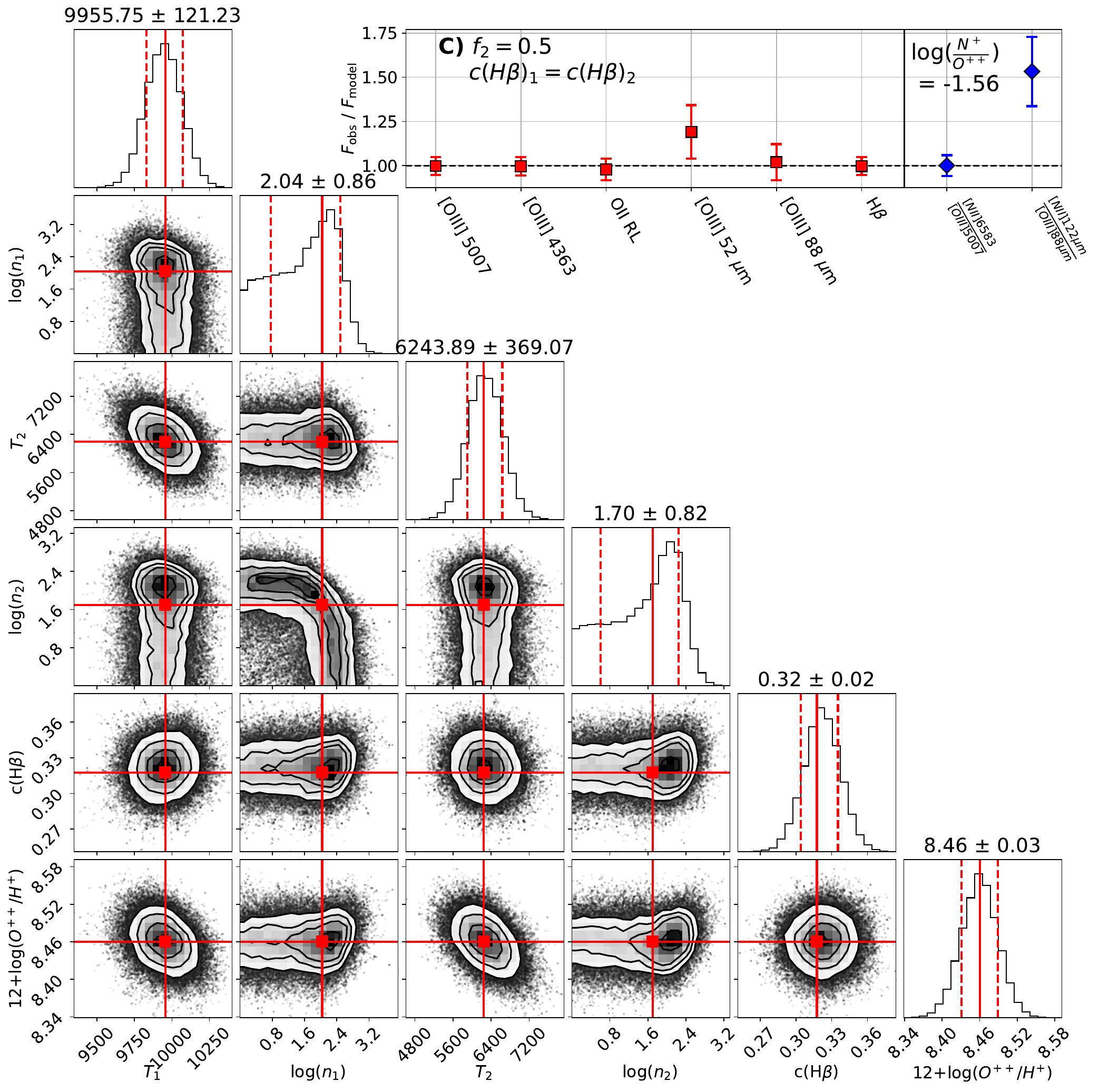}
\caption{ Same as Fig. \ref{fig:corner_a}, but for Model C, which assumes that $T_1, n_1, T_2$, and $n_2$ are independent to each other, but $f_2$ is fixed at 0.5, and $\chb_1 = \chb_2$. 
\label{fig:corner_c}}
\end{figure*}

 Since the model is limited to 6 observational constraints,
we are unable to make all $T_1, T_2, n_1, n_2, \log(O^{++}/H^+), f_2$, and a flux normalizing parameter as free parameters at the same time. However,  to gain insight we consider models with $f_2$ values fixed to those found by Models A and B, namely $f_2=0.17$ and 0.5 for \textit{Model C}. Among these two choices, $f_2 = 0.5$ results in the best fitting result with the lowest $\chi^2$. Fig. \ref{fig:corner_c} shows this result. The fitting result is similar to Model A. The \Oiii\ $\lambda 52~\um$ flux and the \Nii\ $\lambda 122~\um$ / \Oiii\ $\lambda 88~\um$ ratio is also underpredicted by $\sim 1\sigma$. The posterior distributions of $n_1$ and $n_2$ are almost identical. Despite a small difference in the best-fit $n_1$ and $n_2$ values, the difference is within the $1\sigma$ uncertainty, and does not significantly modulate the line fluxes. This further supports the conclusion that non-uniform density distributions are not the main cause of the inconsistent metallicity measurements.  

Models A-C all struggle to match the far-IR line fluxes, suggesting that the far-IR and optical emission may arise from gas in different physical regions.
This is especially evident when comparing the ADF between the far-IR and OCEL metallicities for both $O^{++}/H^+$ and $N^+/H^+$ (see Fig. \ref{fig:metallicities}). The far-IR metallicities are higher than the OCEL metallicities by a similar amount in both $O^{++}/H^+$ and $N^+/H^+$.  In planetary nebulae, dust attenuated high-metallicity inclusions are often invoked to explain the high ($\gtrsim 1$) ADF \citep{liu00, yuan11, danehkar18}. Therefore, we introduce \textit{Model D} which assumes that the two phases have distinctive dust attenuation, $\chb_1$ and $\chb_2$. The combined effect of the two dust phases on the H$\beta$/H$\alpha$ ratio is constrained by the \chb\ posterior measured in \S\ref{sec:attenuation}. The $n_e$ and metallicities are assumed to be the same, and $f_2$ is set at 0.5. Among all models, this model provides the best fit of all the \Oiii\ and \Oiirec\ fluxes. Comparing to Models A and C, this model suggests the differential dust distribution is marginally ($\sim 1\sigma$) required in the system to explain all the O observed line fluxes. The second cooler component has a much higher dust obscuration $\chb_2 = 0.92 \pm 0.28$ than the hotter component typically measured from the OCELs. The cooler component is essentially invisible for the OCEL lines, while the far-IR lines include both the components. Although the \Oiirec\ RLs are similarly affected by dust obscuration compared to the OCEL, they are not affected by the distinctive $T_e$, hence the RL $O^{++} / H^+$ metallicity measured assuming one phase (Fig. \ref{fig:metallicities} and Table \ref{tab:te_met}) is slightly higher from the far-IR compared to the RL. The best-fit $O^{++} / H^+$ from the two-phase model is consistent with the RL metallicity. Despite the success of this model in predicting the fluxes of the O lines, it does not reduce the dicrepancy for the \Nii/\Oiii\ line ratios. 
Therefore, further adjustments are likely required in the two-phase model to explain all the fluxes. We discuss the possibility of these adjustments in the next subsection.  

\begin{figure*}[ht!]
\centering
\includegraphics[width=0.95\textwidth]{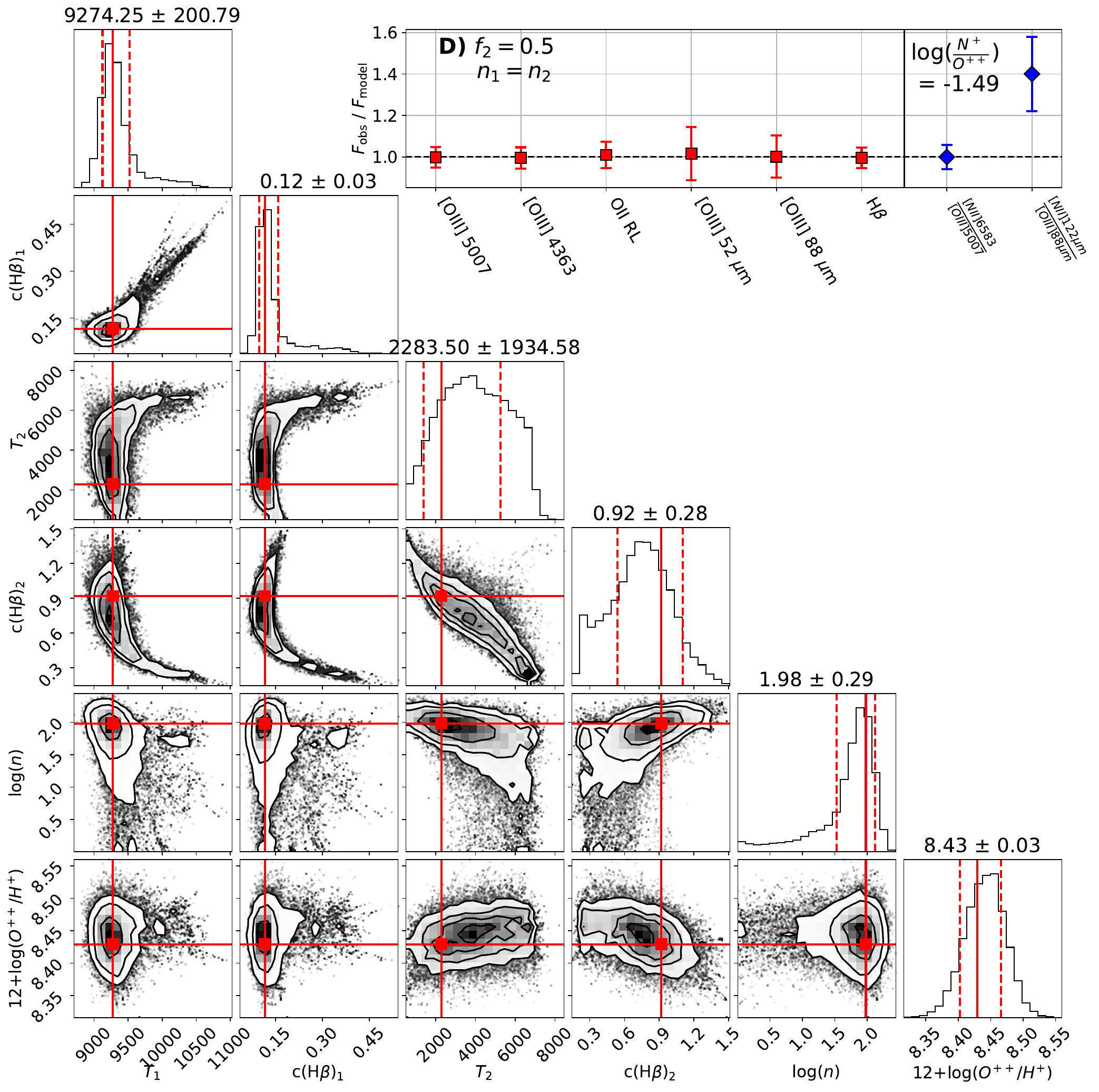}
\caption{ Same as Fig. \ref{fig:corner_a}, but for Model D, which assumes that $T_e$ and \chb\ are distinctive in the two components, but $n_e$ is the same, and $f_2 = 0.5$. 
\label{fig:corner_d}}
\end{figure*}

\subsection{A complex gaseous environment}
\label{sec:complex_gas}

In our analysis, we have established that neither $T_e$ nor $n_e$ fluctuations alone can adequately account for all the observed discrepancies in the optical and far-IR flux ratios of \Oiii\ and \Nii. Notably, the disparities observed in the O$^{++}$/H$^+$ and N$^+$/H$^+$ metallicities, as measured by the OCEL and far-IR methods, suggest that optical photons are fractionally underrepresented. As shown in the two-phase Model D, this scenario can be explained by a portion of gas which is more highly obscured by dust. 
Although we corrected for dust extinction using the available Balmer and Paschen lines, these corrections are biased toward regions of lower dust extinction and can thus underestimate the correction factor. 
Accurate corrections for total attenuation can therefore be a challenge in regions heavily enshrouded by dust.

In Model D, despite having $50\%$ of the gas, the more dust-obscured second component contributes to $\simeq 30\%$ of the total H$\beta$ flux. Meanwhile, the second component contributes to $\simeq 60\%$ of the total far-IR emission line flux. Although we do not fit the \Nii\ / \Oiii\ line ratios in the two-phase models, invoking the dust obscured cooler component can naturally explain the line ratios without invoking an unrealistic amount of temperature fluctuations. The elevated N$^+$/O$^{++}$ ratio identified from far-IR emission relative to the OCEL emission points to higher N/O and/or lower ionization in the more obscured gas. 
Indeed, the mid-IR spectrum shows a $\sim$3$\times$ higher ratio of [\ion{Ne}{2}]~12.8~$\mu$m/\Neiii~15.6~$\mu$m than expected based on optical fluxes \citep{hunt06}, supporting a lower ionization parameter in the more obscured gas.
However, the discrepancy between the RL and OCEL still requires temperature inhomogeneity. 
Model D indeed prefers a significantly lower temperature ($T_e \sim 2000~K$) in the second component, such that it contributes very small OCEL fluxes ($<1 \%$), but $\sim 30\%$ of the RL emission.The presence of this low-$T_e$ component is partially supported by the low $T_e$(\Hei) values (Table~\ref{tab:te_met}), which indicate the existence of a substantially cooler gas component.

This inference regarding dust-obscured gas is further supported by the significant range in dust content estimations reported across various studies. Notably, \citet{johnson04} and \citet{hunt06} have documented that visual extinction ($A_V$) values can vary extensively, ranging from 0.5 to 8, depending on the method of measurement and the spectral region analyzed. Furthermore, \citet{hunt06} employed photoionization modeling of mid-infrared (mid-IR) emission lines and found evidence supporting the concurrent presence of a low-dust optical emitting component alongside a dust-obscured component that is optically invisible. This broad range of inferred dust content highlights the profound influence that obscured gas can exert on our understanding of the chemical composition of \Hii\ regions.
Models incorporating two distinct phases of gas with distinct temperatures have also been posited as an alternative explanation for the ADF compared to models based on a continuous temperature distribution \citep[e.g.,][]{liu00, stasinska02, tsamis03}. In these models, \citet{zhang07} have noted that the inclusion of even a minor fraction ($\sim 10\%$) of low-temperature gas ($\sim 1000$ K) could replicate the observed $\sim 0.2$-dex ADF.

We caution that the results of our analysis of Haro~3 may not be universally applicable, instead highlighting that a range of complex factors can bias the analysis of nebular emission lines.
For example, our conclusions for Haro~3 are at face value contradictory to the results from Mrk~71 reported by \citetalias{chen23}, where we found no significant difference between the far-IR and OCEL O$^{++}$/H$^+$ ratio. However, the major difference between Mrk~71 and Haro~3 is that Haro~3 exhibits substantially ($\gtrsim 3\times$) higher dust attenuation, while Mrk~71 has significantly less dust ($\chb = 0.09 \pm 0.04$), much lower metallicity, and shows no sign of a substantial dust-obscured component.
Our conclusion that density fluctuations do not significantly contribute to the observed abundance discrepancy, especially for the far-IR emission lines, is also somewhat at odds with recent observations from JWST+ALMA of $z>6$ galaxies \citep{harikane25, usui25}. In those works, two-phase models are also introduced to resolve the discrepancies between the OCEL and far-IR \Oiii\ observations, inferring a bimodality with high- and low-density components in those systems. The difference can be explained by the overall high density ($n_e \sim 1000~\mathrm{cm}^{-3}$) and low dust content ($\chb < 0.15$) of those high-$z$ galaxies compared to Haro~3.  
These differences are such that density inhomogeneity can have a much larger effect than dust attenuation in the high-$z$ sources.
Collectively these studies suggest that large fluctuations in $T_e$, $n_e$, dust attenuation, and possibly other properties (e.g., metallicity) can all have significant effects on the observed emission and derived properties, depending on the nature of the sources. Given this complexity, a promising approach to disentangle these effects and establish robust physical properties is to utilize mid-IR features such as the \Siii, \Neiii, and \ion{H}{1} Humphreys lines. These features are relatively insensitive to fluctuations in $T_e$, $n_e$, and attenuation, and can be directly compared with optical CEL measurements. Our group is actively pursuing this avenue for future work.

\section{Summary}

In this study, we conducted a detailed examination of $T_e$ and metallicity from O$^{++}$ and N$^+$ ions using optical and far-IR spectra of Haro~3, a nearby \Hii\ region. The analysis incorporated data from Herschel/PACS, SOFIA/FIFI-LS, and the recently commissioned red channel of Keck/KCWI. The combined capabilities of these three IFU instruments, covering optical and far-IR spectra, enabled precise aperture matching and robust comparisons of absolute fluxes. Our approach has yielded the first abundance discrepancy factor (ADF) measurements for both O$^{++}$ and N$^+$ in a single astronomical object. The principal findings of this research include:

\begin{itemize}
    \item To explain the ADF measured from the \Nii\ far-IR vs. OCEL method, temperature fluctuations would have to be unphysically large. The similar N$^+$/O$^{++}$ ratios observed between the far-IR and OCEL lines indicate that dust obscuration {and the inclusion of cool gas components} are the main causes of the disparity. This is supported by previous work indicating the presence of infrared emission from heavily obscured regions.
    \item We constructed a series of two-phase models to fit the observed flux ratios of \Oiii\ CEL, far-IR, and the \Oiirec\ RLs. These models suggest that the density fluctuations are likely not the leading cause of the observed inconsistencies in those line fluxes. Spatial variations of $n_e$ are present in Haro~3 but are insufficient to significantly modulate the metallicity measurements.  Instead, temperature inhomogeneity and differential dust obscuration are both needed in the best-fit model. 
    \item The differential dust obscuration can also naturally explain the large discrepancies in the OCEL and far-IR \Nii\ / \Oiii\ flux ratios without invoking unphysically large temperature fluctuations. A low $N^+/O^{++}$ ratio is naturally expected in the low-$T_e$ and high-dust component, due to its low ionization and/or high N/O ratio. Such a component would be essentially invisible in the OCEL emission. 
    \item Non-thermal effects, i.e., a $\kappa$ distribution, fail to reconcile the differences between far-IR and OCEL emissions. This model would require a disproportionate injection of high-energy plasma for N$^+$ compared to O$^{++}$, which contradicts their respective ionization levels.
\end{itemize}

Our findings underscore the importance of resolving the origin of the ADF and establishing an accurate metallicity scale. The far-IR fine-structure emission used in this study presents {an innovative tool} for evaluating the {detailed gas properties} in \Hii\ regions, and for cross-validating the results with multiple ions. The combination of far-IR and optical spectra, as utilized here, is especially relevant in light of the strong and auroral line measurements obtained at high redshifts with ALMA and JWST, now reaching $z>8$. 
Our analysis of Haro 3 exemplifies the potential of such measurements to understand the biases inherent in different metallicity measurement techniques, highlighting both their potential and the associated challenges.

However, this study of Haro 3, along with Mrk 71 in \citetalias{chen23}, covers only a limited portion of the parameter space of \Hii\ regions. These cases underline the necessity for a broader investigation to understand the implications for all \Hii\ regions across cosmic history. A comprehensive dataset of \Hii\ regions, characterized by similarly detailed metallicity measurements under a variety of physical conditions, will be essential to this endeavor.
Furthermore, our results also highlight the need for spatially resolved measurements in both optical and IR spectra to reduce the reliance on aperture matching. Future telescopes like PRIMA (The PRobe Far-Infrared Mission for Astrophysics; \citealt{prima23}) would improve the accuracy of similar measurements. Meanwhile, mid-IR observations via JWST provide crucial opportunities to analyze and address the heterogeneity within \Hii\ regions. Notably, mid-IR features such as the H Humphrey lines and the \Siii\ and \Neiii\ lines are relatively unaffected by variations in temperature, density, and dust content, offering more reliable diagnostic capabilities.
Leveraging archival far-IR data combined with JWST mid-IR spectroscopy of nearby \Hii\ regions offers tremendous potential to unravel the underlying causes of the ADF and to refine our understanding of the true metallicity scale, as well as the complex properties of gaseous nebulae. Our team is committed to driving these investigations forward in future studies.

\begin{acknowledgments}
Y.C. and T.J. gratefully acknowledge support for this work from the National Aeronautics and Space Administration
(NASA) under grant 80NSSC23K1132. 
Y.C. is supported by the Direct Grant for Research (C0010-4053720) from the Faculty of Science the Chinese University of Hong Kong.
R.H.-C. thanks the Max Planck Society for support under the Partner Group project ``The Baryon Cycle in Galaxies'' between the Max Planck for Extraterrestrial Physics and the Universidad de Concepci\'on. R.H-C. also gratefully acknowledge financial support from ANID - MILENIO - NCN2024\_112 and ANID BASAL FB210003. The authors would like to thank Karin M. Sandstrom for her contribution to the planning of this work.

This work is based on data obtained at the W. M. Keck Observatory, which is operated as a scientific partnership among the California Institute of Technology, the University of California, and the National Aeronautics and Space Administration. The Observatory was made possible by the generous financial support of the W. M. Keck Foundation. The authors wish to recognize and acknowledge the very significant cultural role and reverence that the summit of Maunakea has always had within the indigenous Hawaiian community.  We are most fortunate to have the opportunity to conduct observations from this mountain. Results in this paper are based on observations made with the NASA/DLR Stratospheric Observatory for Infrared Astronomy (SOFIA). SOFIA is jointly operated by the Universities Space Research Association, Inc. (USRA), under NASA contract NAS2-97001, and the Deutsches SOFIA Institut (DSI) under DLR contract 50-OK-0901 to the University of Stuttgart. Herschel is an ESA space observatory with science instruments provided by European-led Principal Investigator consortia and with important participation from NASA. The authors thank the W. M. Keck Observatory staff, the SOFIA observatory staff, and the Herschel Data Archive for making this study possible. Some pilot studies of this program were conducted on the Palomar 200-inch telescope. We thank the Caltech Optical Observatories for providing the support.

\end{acknowledgments}

\vspace{5mm}
\facilities{Keck(KCWI), SOFIA(FIFI-LS), Herschel(PACS)}
 
\software{astropy \citep{astropy13, astropy18, astropy22}, emcee \citep{emcee13}, PyNeb \citep{luridiana15}, KCWI DRP (\url{https://github.com/Keck-DataReductionPipelines/KcwiDRP}), SOSPEX (\url{https://github.com/darioflute/sospex}), CWITools \citep{cwitools20}, KcwiKit (\url{https://github.com/yuguangchen1/KcwiKit.git})}, and Scientific Colour Maps \citep{crameri18}.

\appendix

\section{Line Fluxes \label{sec:line_fluxes}}

Here, we provide a summary (Table \ref{tab:fluxes}) of the line flux measurements. The fluxes are measured from 1D spectra extracted from Keck/KCWI, SOFIA/FIFI-LS, and Herschel/PACS from two apertures: 1) $r_\mathrm{ap}=2\secpoint25$ for the KCWI PSF, 2) $r_\mathrm{ap}=3"$ after the PSF is matched to the SOFIA/FIFI-LS \Oiii\ $\lambda$52~\um\ data cube, and {3) the attenuation-corrected line instensities that are aperture-matched to the $2\secpoint25$ aperture}. The errors quoted in this table include the pure Poisson noise from observations, and systematic uncertainties from flux calibration [5\% (KCWI) and 11\% (FIFI-LS and PACS)]. Uncertainties from the attenuation correction are additionally included in the dust-corrected intensities. The effect of these uncertainties are considered in all relevant analysis and Tables \ref{tab:te_met} \& \ref{tab:t2} in the paper. For details on how 1D spectra are extracted and fluxes are measured, see \S\ref{sec:flux_measurements}.

\begin{deluxetable}{lcccc}
\tablecaption{{Summary of Haro 3 line fluxes\tablenotemark{a} \label{tab:fluxes}}}
\tablehead{\colhead{Line} & $\lambda_0$\tablenotemark{b} & $F(r_\mathrm{ap}$ = 2\secpoint25) & $F(r_\mathrm{ap} = 3")$ & $I(r_\mathrm{ap} = 2\secpoint25)$\tablenotemark{c}\\
& $(\mathrm{\AA})$ & \multicolumn{2}{c}{($10^{-16}~\mathrm{erg~s}^{-1}~\mathrm{cm}^{-2}$)} & ($I(H\beta) = 100$) }
\startdata
\Oii\ & 3726.03 & $2010 \pm 110$ & $909 \pm 48$ & $93.3 \pm 6.8$ \\
\Oii\ & 3728.82 & $2550 \pm 140$ & $1224 \pm 63$ & $118.4 \pm 8.4$ \\
H10 & 3797.90 & $90.3 \pm 5.1$ & $19.9 \pm 2.0$ & $4.15 \pm 0.30$\\
H9  & 3835.39 & $151.9 \pm 8.0$ & $43.9 \pm 2.8$ & $6.95 \pm 0.49$ \\
\Neiii & 3868.75 & $602 \pm 30$ & $208 \pm 11$ & $27.4 \pm 1.9$ \\
{H8+\ion{He}{1}}  & 3889.05 & $417 \pm 21$ & $142.7 \pm 7.3$ & $18.9 \pm 1.3$ \\
H7+\Neiii  & 3970.07 & $522 \pm 26$ & $148.3 \pm 7.6$ & $23.3 \pm 1.6$ \\
\Hei & 4009.22 & $2.85 \pm 0.26$ & --- & $0.126 \pm 0.013$ \\
\Hei & 4026.21 & $41.5 \pm 2.1$ & $16.16 \pm 0.90$ & $1.83 \pm  0.12$ \\
\Sii & 4068.60 & $33.4 \pm 1.7$ & $16.94 \pm 0.96$ & $1.461 \pm 0.097$ \\
\Sii & 4076.35 & $9.94 \pm 0.55$ & --- & $0.434 \pm 0.030$ \\
H$\delta$ & 4101.74 & $600 \pm 30$ & $209 \pm 11$ & $27.2 \pm 1.7$\\
\Hei & 4120.82 & $2.66 \pm 0.22$ & $3.18 \pm 0.33$ & $0.115 \pm 0.010$ \\
\Hei & 4143.76 & $3.68 \pm 0.22$ & --- & $0.159 \pm 0.011$ \\
H$\gamma$ & 4340.47 & $1153 \pm 58$ & $424 \pm 21$ & $47.5 \pm 3.1$ \\
\Oiii\   & 4363.21 &  $52.4 \pm 2.7$ & $19.5 \pm 1.3$ & $2.15 \pm 0.14$ \\
\Hei & 4387.93 & $10.90 \pm 0.59$ & $5.45 \pm 0.51$ & $0.444 \pm 0.030$ \\
\Hei & 4471.09 & $103.8 \pm 5.2$ & $36.6 \pm 1.9$ & $4.15 \pm 0.27$ \\
\Oiirec\ & 4638.86 &  $1.76 \pm 0.17$ & --- & $0.0678 \pm 0.0069$ \\
\Oiirec\ & 4641.81 &  $2.45 \pm 0.19$ & --- & $0.0943 \pm 0.0081$ \\
\Oiirec\ & 4649.13 &  $0.82 \pm 0.22$ & --- & $0.0315 \pm 0.0085$ \\
\Oiirec\ & 4650.84 &  $1.17 \pm 0.18$ & --- & $0.0449 \pm 0.0073$ \\
\Oiirec\ & 4661.63 &  $2.36 \pm 0.18$ & --- & $0.0904 \pm 0.0078$ \\
\Oiirec\ & 4676.23 &  $0.78 \pm 0.15$ & --- & $0.0298 \pm 0.0060$ \\
\Ariv\   & 4711.37 &  $4.02 \pm 0.23$ & $1.92 \pm 0.17$ & $0.152 \pm 0.011$ \\
\Hei & 4713.14 & $10.29 \pm 0.53$ & $3.65 \pm 0.26$ & $0.390 \pm 0.025$ \\
\Ariv\   & 4740.16 &  $3.06 \pm 0.20$ & $1.02 \pm 0.31$ & $0.1153 \pm 0.0087$ \\
H$\beta$ & 4861.33 &  $2720 \pm 140$ & $1004 \pm 50$ & $100.0 \pm 6.1$ \\
\Hei & 4921.93 & $22.5 \pm 1.3$ & --- & $0.818 \pm 0.051$ \\
\Oiii\   & 4958.91 &  $3650 \pm 180$ & $1186 \pm 59$ & $131.7 \pm 7.9$ \\
\Oiii\   & 5006.84 &  $11000 \pm 550$ & $3570 \pm 179$ & $392 \pm 24$ \\
\Hei & 5015.68 & $58.1 \pm 8.7$ & $18.7 \pm 3.1$ & $2.07 \pm 0.32$ \\
\Cliii\  & 5517.71 &  $15.3 \pm 1.2$ & $8.69 \pm 0.92$ & $0.505 \pm 0.044$ \\
\Cliii\  & 5537.88 &  $11.4 \pm 1.2$ & $6.75 \pm 0.91$ & $0.376 \pm 0.042$ \\
\enddata
\tablenotemark{a}{All uncertainties contain the flux-calibration uncertainty (5\% for the optical, and 11\% for the far-IR emission lines).}
\tablenotetext{b}{Rest-frame air wavelengths.}
\tablenotetext{c}{Dust-corrected intensity, aperture-matched to the $2\secpoint25$ aperture. Including both the flux calibration and dust correction uncertainties.}
\tablenotetext{d}{Measured simultaenously with fixed \Nii\ $\lambda\lambda$6583/6548 ratio of 2.942.}
\tablenotetext{e}{Total flux of the \Oii\ $\lambda\lambda$7318.39, 7319.99 lines.}
\tablenotetext{f}{Total flux of the \Oii\ $\lambda\lambda$7329.66, 7330.73 lines.}
\end{deluxetable}

\setcounter{table}{3}
\begin{deluxetable}{lcccc}
\tablecaption{---\textit{Continued}}
\tablehead{\colhead{Line} & $\lambda_0$\tablenotemark{b} & $F(r_\mathrm{ap}$ = 2\secpoint25) & $F(r_\mathrm{ap} = 3")$ & $I(r_\mathrm{ap} = 2\secpoint25)$\tablenotemark{c}\\
& $(\mathrm{\AA})$ & \multicolumn{2}{c}{($10^{-16}~\mathrm{erg~s}^{-1}~\mathrm{cm}^{-2}$)} & ($I(H\beta) = 100$) }
\startdata
\Nii\    & 5754.64 &  $8.90 \pm 0.50$ & $2.57 \pm 0.49$ & $0.286 \pm 0.018$ \\
\Hei & 5875.64 & $397 \pm 20$ & $141.1 \pm 7.2$ & $12.56 \pm 0.72$ \\
\Siii & 6312.10 & $44.0 \pm 2.2$ & $14.26 \pm 0.84$ & $1.333 \pm 0.078$ \\
\Nii{}\tablenotemark{d}   & 6548.03 &  $233 \pm 12$ & $103.7 \pm 5.3$ & $6.89 \pm 0.39$ \\
H$\alpha$ & 6562.82 & $10120 \pm 510$ & $3610 \pm 180$ & $299 \pm 17$ \\ 
\Nii{}\tablenotemark{d}    & 6583.41 &  $685 \pm 35$ & $305 \pm 16$ & $20.2 \pm  1.1$ \\
\Hei & 6678.15 & $124.8 \pm 6.3$ & $43.9 \pm 2.2$ & $3.65 \pm 0.20$ \\
\Sii\    & 6716.47 &  $610 \pm 31$ & $305 \pm 15$ & $17.76 \pm 1.0$ \\
\Sii\    & 6730.85 &  $491 \pm 25$ & $233 \pm 12$ & $14.27 \pm 0.80$ \\
\Hei & 7065.28 & $105.6 \pm 5.3$ & $32.8 \pm 1.7$ & $2.97 \pm 0.17$ \\
\Ariii & 7135.78 & $353 \pm 18$ & $122.2 \pm 6.1$ & $9.87 \pm 0.54$ \\
\Hei & 7281.35 & $18.69 \pm 0.99$ & $2.96 \pm 0.45$ & $0.515 \pm 0.030$ \\
\Oii & 7319\tablenotemark{e} & $97.5 \pm 4.9$ & $29.4 \pm 1.5$ & $2.68 \pm 0.15$ \\
\Oii & 7330\tablenotemark{f} & $83.7 \pm 4.2$ & $30.1 \pm 1.5$ & $2.30 \pm 0.13$ \\
\Ariii & 7751.10 & $89.4 \pm 4.5$ & $24.7 \pm 1.6$ & $2.35 \pm 0.12$ \\
P14      & 8598.39 &  $29.4 \pm 1.5$ & $11.25 \pm 0.58$ & $0.721 \pm 0.040$ \\
P10      & 9014.91 &  $80.3 \pm 4.0$ & $22.0 \pm 1.2$ & $1.92 \pm 0.10$ \\
\Siii & 9068.90 & $1108 \pm 57$ & $313 \pm 16$ & $26.4 \pm 1.5$ \\\hline
\Oiii{}  & 51.8145~\um\ & --- & $3460 \pm 470$ & $188 \pm 24$ \\
\Oiii{}  & 88.3564~\um\ & --- & $4200 \pm 460$ & $225 \pm 25$ \\
\Nii{}   & 121.769~\um\ & --- & $41.9 \pm 5.7$ & $2.27 \pm 0.29$ \\
\enddata
\end{deluxetable}

%% For this sample we use BibTeX plus aasjournals.bst to generate the
%% the bibliography. The sample631.bib file was populated from ADS. To
%% get the citations to show in the compiled file do the following:
%%
%% pdflatex sample631.tex
%% bibtext sample631
%% pdflatex sample631.tex
%% pdflatex sample631.tex

\bibliography{sample631}{}
\bibliographystyle{aasjournal}

%% This command is needed to show the entire author+affiliation list when
%% the collaboration and author truncation commands are used.  It has to
%% go at the end of the manuscript.
%\allauthors

%% Include this line if you are using the \added, \replaced, \deleted
%% commands to see a summary list of all changes at the end of the article.
%\listofchanges

\end{CJK*}
\end{document}